\documentclass[aps,prb,reprint,superscriptaddress]{revtex4-2}

\usepackage[T1]{fontenc}
\usepackage[utf8]{inputenc}

\usepackage{amsmath,amssymb,graphicx,bm}
\usepackage[caption=false]{subfig}
\usepackage{xcolor}

\newcommand{\br}{\bm{r}}
\newcommand{\qv}{\bm{q}}
\newcommand{\suc}{_{\mathrm{c}}}
\newcommand{\sxc}{_{\mathrm{xc}}}
\newcommand{\fxc}{f\sxc}
\newcommand{\re}{\mathrm{Re}~}
\newcommand{\im}{\mathrm{Im}~}
\newcommand{\kf}{k_{\mathrm{F}}}
\newcommand{\rs}{r_{\mathrm{s}}}
\newcommand{\xks}{\chi_0(\qv,\omega)}
\newcommand{\kt}{\widetilde{k}}
\newcommand{\veff}{v_{\mathrm{eff}}}
\newcommand{\vb}{v_{\mathrm{bare}}}

\begin{document}

\title{First-principles wavevector- and frequency-dependent exchange-correlation kernel for jellium at all densities}

\author{Aaron D. Kaplan}
\email{kaplan@temple.edu}
\affiliation{Department of Physics, Temple University, Philadelphia, PA 19122}
\author{Niraj K. Nepal}
\affiliation{Department of Physics, Temple University, Philadelphia, PA 19122}
\author{Adrienn Ruzsinszky}
\affiliation{Department of Physics, Temple University, Philadelphia, PA 19122}
\author{Pietro Ballone}
\affiliation{School of Physics and Conway Institute for Biomolecular and Biomedical Research, University College, Dublin, Ireland}
\author{John P. Perdew}
\email{perdew@temple.edu}
\affiliation{Department of Physics, Temple University, Philadelphia, PA 19122}
\affiliation{Department of Chemistry, Temple University, Philadelphia, PA 19122}

\date{\today}

\begin{abstract}
  We propose a spatially and temporally nonlocal exchange-correlation (xc) kernel for the spin-unpolarized fluid phase of ground-state jellium, for use in time-dependent density functional and linear response calculations. The kernel is constructed to satisfy known properties of the exact xc kernel, to accurately describe the correlation energies of bulk jellium, and to satisfy frequency-moment sum rules at a wide range of bulk jellium densities, including those low densities that display strong correlation and symmetry breaking. These effects are easier to understand in the simple jellium model than in real systems. All exact constraints satisfied by the recent MCP07 kernel [A. Ruzsinszky, \textit{et al.}, Phys. Rev. B \textbf{101}, 245135 (2020)] are maintained in the new revised MCP07 (rMCP07) kernel, while others are added.
  The revision $\fxc^\text{rMCP07}(q,\omega)$ differs from MCP07 only for non-zero frequencies $\omega$.
  Only at densities much lower than those of real bulk metals is the frequency dependence of the kernel important for the correlation energy of jellium.
  As the wavevector $q$ tends to zero, the kernel has a $-4\pi \alpha(\omega)/q^2$ divergence whose frequency-dependent ultranonlocality coefficient $\alpha(\omega)$ vanishes in jellium, and is predicted by rMCP07 to be extremely small for the real metals Al and Na.
\end{abstract}

\maketitle

\section{Introduction}

Ground-state density functional theory (g.s. DFT) \cite{kohn1965} is a mature field that yields exact-in-principle ground-state energies and densities of any non-relativistic many-electron system. Practical applications of g.s. DFT require approximations to the ``exchange-correlation'' energy $E\sxc$, the simplest of which, the local density approximation (LDA), predates modern g.s. DFT. Modern approximations to the xc energy can make reasonable predictions of g.s. properties, often comparable to experiment.

Ground state DFT can be extended to the time domain to include either arbitrary \cite{runge1984} or weak \cite{gross1985,petersilka1996} time-dependent external potentials (TD-DFT). Within the exact theory or the linear-response regime, the xc potential rather than the xc energy must be approximated. The xc kernel $\fxc$ is related to the exchange correlation potential $v\sxc$ via functional differentiation
\begin{equation}
  f\sxc(\br,t;\br',t') = \frac{\delta v\sxc(\br,t)}{\delta n(\br',t')} \theta(t-t')
\end{equation}
with $\theta(y>0)= 1$, and $\theta(y<0) = 0$. $\fxc$ can be computed from the second functional derivative of $E\sxc$ from a g.s. calculation only in an adiabatic approximation (assuming the response is local in time). Approximate expressions for $E\sxc$ used in g.s. calculations do not necessarily provide similarly accurate adiabatic approximations to $f\sxc$ for use in TD-DFT calculations.

Thus, highly-accurate approximations to the exact $f\sxc$ are needed for realistic beyond-RPA descriptions of materials. G.S. DFT is instructive in this regard: functionals that are most broadly transferrable, e.g., that of Ref. \cite{sun2015}, are designed to satisfy known limiting behaviors of the exact $E\sxc$. These include the uniform density (jellium) limit, gradient expansions for slowly-varying metallic densities, and scaling relations. Being able to find $E\sxc$ (or $\fxc$) for the simple jellium model is necessary but insufficient for computation of $E\sxc$ (or $\fxc$) in real materials.

Recently, an approximate, dynamic kernel for jellium was proposed with similar construction principles. Jellium is characterized by a uniform electron density $n = 3/(4\pi \rs^3) = \kf^3/(3\pi^2)$. In this work, we will use Hartree atomic units, $\hbar = m_e = e^2 = 1$, for all quantities and numerical coefficients, unless noted otherwise. The modified Constantin-Pitarke 2007 (MCP07) \cite{ruzsinszky2020} kernel is constructed as an interpolation between static $\fxc(q,\omega=0)$ and long-wavelength dynamic $f\sxc(q=0,\omega)$ limits,
\begin{align}
  \fxc^{\mathrm{MCP07}}(q,\omega) = &\left\{1 + e^{-k q^2}\left[\frac{\fxc(0,\omega)}{\fxc(0,0)} - 1\right] \right\}\nonumber \\
  & \times \fxc^{\mathrm{MCP07}}(q,0). \label{eq:mcp07}
\end{align}
In this equation, $\fxc(0,\omega)$ is the Gross-Kohn-Iwamoto (GKI) kernel \cite{gross1985,iwamoto1987}, which satisfies known analytic and asymptotic $\omega\to\infty$ behaviors of the exact $\fxc(0,\omega)$. The static limit is controlled by $\fxc^{\mathrm{MCP07}}(q,0)$, a revision to the Constantin-Pitarke static kernel \cite{constantin2007} that enforces known exact constraints on the short-wavelength limit $\fxc(q\to\infty,0)$, as well as the gradient expansion of $\fxc(q,0)$ for slowly-varying densities. $\fxc(0,0)$ is the adiabatic local density approximation (ALDA), found as the $q\to 0$ limit of the Fourier transform of $\frac{\delta^2 E\sxc}{\delta n(\br)\delta n(\br')}$ evaluated at the uniform density $n$. The order in which the $|\bm{q}|\to 0$ and $\omega \to 0$ limits are taken yields different limiting behaviors for the exact $\fxc$, as discussed in Appendix C. For MCP07 and our model $\fxc$, we make the simplifying approximation that either order of limits yields the ALDA $\fxc$. The inverse-squared screening wavevector
\begin{equation}
    k = -\frac{\fxc(0,0)}{4\pi B(\rs)} \label{eq:k_mcp07}
\end{equation}
with $B(\rs)$ parameterized by Eq. 7 of Ref. \cite{corradini1998}, was chosen to enforce two separate exact constraints on the static kernel $\fxc(q,\omega=0)$ \cite{ruzsinszky2020}
\begin{align}
    \lim_{q\to 0} \left[ \lim_{\omega \to 0} \fxc(q,\omega)\right] &= \fxc(0,0) \\
    \lim_{q\to \infty} \left[ \lim_{\omega \to 0} \fxc(q,\omega)\right] &= -4\pi\left[\frac{C(\rs)}{\kf^2} + \frac{B(\rs)}{q^2} \right].
\end{align}
$C(\rs)$ is given by Eq. A2 of Ref. \cite{constantin2007}. However, $k$ also appears, through \(e^{-k q^2}\), in the dynamic MCP07 to control the interpolation in Eq. (\ref{eq:mcp07}) between the non-uniform static and uniform dynamic limits. This choice was made consistent with an Occam's razor principle: Other things being equal, the
simplest hypothesis is to be preferred. We will investigate the effect of modifying $k$ in $e^{-k q^2}$.

It should be kept in mind that the random phase approximation (RPA), which sets $\fxc^{\text{RPA}} = 0$, includes exchange effects and long-range correlation effects exactly in metals \cite{langreth1977}. The RPA lacks an accurate description of short-range correlation \cite{singwi1968}, which is typically better described by semi-local g.s. energy \emph{functionals} (depending only upon the electron density and its spatial derivatives), motivating the family of RPA+ energy functionals \cite{kurth1999}. These can provide highly-accurate descriptions of metals, but do not test $\fxc$. In RPA+, a local or semi-local correction is added to RPA.

Although the ALDA, by definition, provides a better description of short-range correlation than does the RPA, ALDA does not generally make better predictions than RPA. This can be seen clearly in Fig. S10 of Ref. \cite{perdew2021} which plots jellium correlation energies per electron $\varepsilon\suc$: the RPA makes $\varepsilon\suc$ too negative, whereas the ALDA over-corrects RPA at all densities. The ALDA also predicts onset of a static charge density wave for $\rs \approx 30$, not in line with any quantum Monte Carlo (QMC) predictions of Wigner crystallization. A transition from the spin-unpolarized fluid phase to the Wigner crystal phase is possible for $\rs \approx 85 \pm 20$ bohr \cite{ceperley1980}.

It should be noted that the exact value of $\rs$ for which the fermion fluid crystallizes in jellium is still uncertain.
The earliest reliable prediction of a transition from the \textit{ferromagnetic} fluid phase to the Wigner crystal phase from QMC was $\rs = 100 \pm 20$ bohr \cite{ceperley1980}, with more recent QMC calculations finding $\rs = 65 \pm 10$ bohr \cite{ortiz1999} and $\rs=106 \pm 1$ bohr \cite{drummond2004}.
As the energy differences separating the Wigner crystal and fluid phases of low-density jellium are extremely small (on the order of $10^{-4}$--$10^{-5}$ eV \cite{ceperley1980}), any small numerical, methodological, etc. errors can drastically alter the predicted phase diagram at low densities, including the relative ordering of the fluid phases.
Moreover, each of the references cited here used different approximation methods, and different methods to estimate the uncertainty in their results.
This makes a direct comparison nontrivial.

For the present purposes of this work, however, it suffices to know that: (1) the Wigner crystallization phase is energetically competitive with the fluid phases for jellium at densities $\rs \geq 60$; (2) the structure factor of the fluid phase is very weakly spin-dependent at these densities \cite{holzmann2020}.
Neither observation depends upon the precise values given previously, but both are relevant for the construction of the kernel presented here.

Extensive tests of the MCP07 functional for real systems are not currently available, and not without good reason, as we shall discuss shortly.
However, it was observed in Ref. \cite{perdew2021} that the MCP07 kernel can be improved in two regards: a more accurate recovery of jellium correlation energies at all densities, and better satisfaction of the third frequency-moment sum rule (see, for example, Eq. 3.142 of Ref. \cite{giuliani2005}) for low-density jellium.
Although the densities at which the MCP07 correlation energy is seriously in error are too low to be important in real materials, they are the densities at which jellium displays the interesting effects of strong correlation and symmetry breaking. These effects are easier to understand in a simple model like jellium than they are in real materials.
This motivates the main inquiry of this paper: improving the MCP07 kernel for jellium at all densities and for known exact sum rules.

Applications of the unmodified MCP07 and rMCP07 kernels to real systems are likely to be limited to metals.
Intermetallic formation energies are described rather poorly by RPA, but improve somewhat \cite{nepal2020} with a wavevector-dependent uniform gas kernel, and might improve further with the MCP07 or rMCP07 kernels.

\section{Comparing CP07, MCP07, and a novel model kernel \label{sec:fxc_comp}}

The construction principles underlying CP07 are the common link between all three kernels, although each differs substantially in their wavevector and frequency dependence.
In analogy with g.s. DFT \cite{sun2015}, we refer to their common construction principle as the satisfaction of exact constraints.
One constructs an approximate kernel by interpolating between known limits of the exact $\fxc$ for jellium.
The exact constraints imposed on MCP07 seem to suffice only for the density range $\rs \leq 10$ bohr, which includes the typical range of  electron densities in metals.
This range is of obvious importance for practical purposes.
We will argue that a good deal of interesting physics is contained in the less-studied, lower-density jellium.

The CP07 kernel is constructed for wavevectors $q$ and imaginary frequencies $\omega = i u$ only, \cite{constantin2007}
\begin{align}
    \fxc^\text{CP07}(q,u) &= \frac{4\pi}{q^2} B(\rs)\{ \exp[-K(\rs,u) q^2] - 1\} \nonumber \\
    & - \frac{4\pi}{\kf^2}\frac{C(\rs)}{1 + 1/q^2}. \label{eq:cp07}
\end{align}
The $B(\rs)$ function is given by Eq. (7) of Ref. \cite{corradini1998}, and the $C(\rs)$ function is given by Eq. (A2) of Ref. \cite{constantin2007}.
All frequency dependence is contained within the function $K(\rs,u)$; to evaluate the kernel at real frequencies (or at arbitrary complex frequencies), one must find the analytic continuation of the kernel.
As noted in Ref. \cite{ruzsinszky2020}, the approach to the large-$q$ limit of CP07 is not quite right.
To compensate for that, the CP07 $K(\rs,u)$ is fitted to ensure that $\fxc^\text{CP07}$ reproduces the correlation energies per electron found with the Perdew-Wang \cite{perdew1992} local spin-density approximation (LSDA).
$K(\rs,u)$ is a rational polynomial in $u$.

MCP07 builds upon CP07 in a few substantial ways:
\begin{enumerate}
    \item introducing an interpolation between zero and infinite frequency limits, allowing for a more-controlled frequency dependence;
    \item using a function of real-valued frequency that is easily continued to complex frequencies;
    \item correcting CP07's approach to the $q\to\infty$ limit;
    \item making the gradient expansion coefficients for weakly-inhomogeneous densities more accurate (small $q$ regime).
\end{enumerate}
MCP07 adopts the structure of CP07 only for its static limit, modifying the screening wavevector to have only density-dependence, \cite{ruzsinszky2020}
\begin{align}
    \fxc^\text{MCP07}(q,0) &= \frac{4\pi}{q^2} B(\rs)\{ \exp[-k(\rs) q^2](1 + E(\rs) q^4) - 1\} \nonumber \\
    & - \frac{4\pi}{\kf^2}\frac{C(\rs)}{1 + 1/(k q^2)^2}.
\end{align}
\(E(\rs)\), defined in Eq. (14) of Ref. \cite{ruzsinszky2020}, controls the second-order gradient expansion, and $k(\rs)$, shown in Eq. (\ref{eq:k_mcp07}), ensures recovery of the ALDA when $q\to0$.
By correcting the wavevector dependence, including the correct second-order gradient expansion omitted in CP07, MCP07 is able to predict both the emergence of a static charge-density wave in low-density jellium, and a transition density in the correct range; CP07 does not predict onset of a static charge-density wave \cite{ruzsinszky2020}.

The MCP07 model has no fitted parameters, but predicts accurate correlation energies for jellium in a metallic range of densities.
The \textit{static} MCP07 kernel is also highly-accurate in its predictions of jellium correlation energies.
This observation confirms the conjecture of Lein, Gross and Perdew \cite{lein2000} that the correlation energies of high- and metallic-density jellium are largely determined by the wavevector-dependence of the kernel, and are much less sensitive to its frequency-dependence.
They advanced this argument after noticing that the Richardson-Ashcroft kernel \cite{richardson1994} and its static limit predicted similarly accurate correlation energies at higher densities.
Recently, this conjecture was confirmed \cite{woods2021} in finite one-dimensional systems by comparing the energies computed using the exact kernel and its static limit.
As we will show, this conjecture does not apply at lower densities (in three dimensions).

The frequency-dependence of the MCP07 kernel, controlled by $\fxc(0,\omega)$ separately from the static kernel $\fxc^\text{MCP07}(q,0)$, is modeled by the Gross-Kohn \cite{gross1985} dynamic local density approximation (LDA), with a correct high frequency limit due to Iwamoto and Gross \cite{iwamoto1987}.
We hereafter refer to this kernel as the GKI dynamic LDA. In CP07, the frequency dependence was chosen to satisfy first and third moment frequency sum rules (Eqs. 3.141 and 3.142 of Ref. \cite{giuliani2005}) in the $q\to0 $ limit.
(Ref. \cite{perdew2021} demonstrates that a dynamic kernel satisfying the third-frequency moment sum rule in this limit does not necessarily satisfy it for all $q$.)
The GKI dynamic LDA is constructed for real frequencies, and satisfies the same sum rules as CP07.
It is easily continued to arbitrary complex frequencies.

To better emphasize the construction principles underlying the new XC kernel presented here, we refer to this new kernel as the revised MCP07 (rMCP07) kernel.
rMCP07 retains all exact constraints satisfied by CP07 and MCP07, and adds a few auxiliary constraints: accurate description of the jellium structure factor, sum rules, and correlation energies at all densities.
These constraints were already satisfied sufficiently by MCP07 in the typical metallic range of densities, but not at lower densities \cite{perdew2021}.

By design, rMCP07 makes modest corrections to MCP07 in the metallic range of densities, and more substantial corrections in the intermediate-to-low range of densities.
For practical purposes, this means that rMCP07 and MCP07 should be comparably accurate for typical metals - although rMCP07 also prescribes a numeric parameterization of the analytic continuation of the kernel to imaginary frequencies, a boon for computational efficiency.

From a theoretical standpoint, low-density jellium models exotic phenomena that are often associated with complex materials: strong correlation \cite{wigner1934,seidl2007} and symmetry-breaking \cite{ceperley1980,ortiz1994,perdew2021}, among others.
An accurate model of $\fxc$ at low densities is needed to further study emergent phenomena in jellium.
Because jellium is simple in comparison to real systems, the origins of these effects can be most easily understood in the jellium model.
Both MCP07 and rMCP07 correctly predict a drop in the spectral function towards zero frequency around the known wavevector of the incipient static charge-density wave, as shown in Ref. \cite{perdew2021} and here.

In g.s. DFT, the LSDA is the uniform-density limit of more sophisticated approximations to the XC energy (e.g., Ref. \cite{sun2015}).
LSDA is constructed to accurately model the XC energy of jellium at all physical spin-densities.
XC energy functionals that tend to the LSDA for uniform densities have been shown to describe $sp$-bonded molecules more accurately than those that do not \cite{zope2019}.
These systems are completely dissimilar to jellium, but still have energetically-relevant regions of lower inhomogeneity that are well-described by LSDA.

In the same way, construction of general-purpose kernels for real materials should be aided by construction of a highly-accurate, approximate kernel for jellium, where the $q \to 0$ limit of the kernel is a finite negative number.
We do not suggest that a kernel for jellium can accurately describe systems like insulators, for which it was determined empirically that the correct long-wavelength limit of the kernel is \cite{reining2002}
\begin{equation}
    \lim_{q\to 0} \fxc(\bm{q},\bm{q},\omega) = -\frac{4\pi \alpha(\omega)}{q^2}.
\end{equation}
The functional form of $\alpha(\omega)$, often called the ``ultranonlocality'' coefficient, is not known in general.
Empirical approximations using material-specific parameters (e.g., Ref. \cite{botti2005}) typically use either experimental data or results from higher-level theories to fit a model for $\alpha(\omega)$.
Appendix \ref{sec:app_unonloc} presents approximate values of this coefficient in metals, calculated from a formula for weakly-inhomogeneous systems using the jellium kernel developed here.
Many empirical kernels for real systems model this behavior, but they contain parameters that are fitted to experimental data or g.s. DFT input.
A general purpose construction would not rely (so heavily) on empiricism.
Determining an accurate, approximate kernel for jellium is a necessary but insufficient step for constructing a general-purpose kernel for real materials, including metals.

We will demonstrate the versatility of this kernel by calculating physical quantities that have interpretations in real systems, and not with self-consistent calculations. A few freely available codes, e.g., GPAW \cite{gpaw_tddft} and the DP code \cite{dp_code}, can perform self-consistent TD-DFT calculations in solids using a model $\fxc(q,\omega)$ as input. However, obtaining well-converged solutions in real systems is often extremely challenging, and deserves due attention in a dedicated computational work. As this is beyond the scope of the current work, we will instead focus on direct applications of the rMCP07 kernel to physical properties, such as screening due to a weak perturbation. As another direct application of our kernel, one could use Eqs. 21 and 23 of Ref. \cite{vanzini2021} to construct a fully nonlocal approximation to the exchange-correlation potential for a given density.

There are practical limitations to using a model $\fxc[n](q,\omega)$ in TD-DFT codes. If, for all real frequencies, only the imaginary part of the kernel is defined in closed form, the real part must be computed by a Kramers-Kronig relation. If the kernel is defined in closed form only at real frequency, one must then analytically continue the kernel to imaginary frequencies to efficiently compute correlation energies, as will be discussed. The continuation is typically done by numeric integration, or Taylor expansion. The cost of repeated numeric integration (or series expansion) compounds substantially. Our solutions to these problems will be discussed in Section \ref{sec:new_model}.

\section{Revised MCP07 XC kernel: \MakeLowercase{r}MCP07 \label{sec:new_model}}

We begin by re-parameterizing $\re \fxc(0,\omega)$ at real frequencies $\omega$. Note that the Gross-Kohn-Iwamoto kernel proposes only an imaginary part of $\fxc(q=0,\omega)$, and the real part must be constructed via the Kramers-Kronig relation
\begin{equation}
  \re \fxc(0,\omega) - \fxc(0,\infty) = \frac{1}{\pi}\mathrm{P}\int_{-\infty}^{\infty} \frac{\im \fxc(0,u)}{u - \omega} du.
\end{equation}
Iwamoto and Gross determined the infinite-frequency limit to be \cite{iwamoto1987}
\begin{equation}
  \fxc(0,\infty) = -\frac{1}{5} \frac{3\pi}{\kf^2} - \frac{1}{15 n}\left[22 \varepsilon^{\mathrm{UEG}}\suc + 26 \rs \frac{d \varepsilon^{\mathrm{UEG}}\suc}{d \rs}\right],
\end{equation}
with $\varepsilon^{\mathrm{UEG}}\suc$ the correlation energy per electron in a uniform electron gas (UEG). Reference \cite{perdew2021} determined that the frequency-dependence of the MCP07 kernel at ``intermediate'' $\rs$ (particularly $\rs = 69$) was likely in error, as the static structure factor
\begin{equation}
  S(\qv) = \int_{0}^{\infty}S(\qv,\omega) d\omega \label{eq:sq_def}
\end{equation}
exhibited unphysically large peaks \cite{priv_comm}, as compared to previously unpublished QMC data \cite{ortiz1999} shown in Fig. \ref{fig:qmc_dat} of Appendix B.
Here, we define the term ``intermediate'' densities as that range of densities between normal metallic densities ($1 \lesssim \rs \lesssim 10$) and the Wigner crystal phase of jellium ($\rs \gtrsim 85$).
Thus we will use ``intermediate density'' to refer to the approximate range $10 \lesssim \rs \lesssim 100$.
The dynamic structure factor, or spectral function,
\begin{equation}
  S(\qv,\omega) = -\frac{1}{\pi n}\im \chi(\qv,\omega)
\end{equation}
is determined by the adiabatic-connection fluctuation-dissipation theorem \cite{nozieres1958,langreth1975} for the interacting density-density response function
\begin{equation}
  \chi(\qv,\omega) = \frac{\xks}{1 - [4\pi/\qv^2 + \fxc(\qv,\omega)]\xks}, \label{eq:chi_recip}
\end{equation}
and $\xks$ is the non-interacting, or Kohn-Sham, response function \cite{lindhard1954}.

In the MCP07 kernel, $\re \fxc(0,\omega)$ is parametrized as
\begin{align}
  \re \fxc(0,\omega) &= \fxc(0,\infty) - c [b(n)]^{3/4}h(\widetilde{\omega}), \label{eq:re_fxc_gki} \\
  \widetilde{\omega} &= [b(n)]^{1/2} \omega \\
  b(n) &= \left\{\frac{\gamma}{c}[\fxc(0,\infty)-\fxc(0,0)]\right\}^{4/3}
\end{align}
where $\gamma = \Gamma(\frac{1}{4})^2/(32\pi)^{1/2}$, and $c = 23\pi/15$ are determined from the static and infinite frequency limits of \cite{gross1985}
\begin{align}
    \im \fxc(0,\omega) &= - c [b(n)]^{3/4} g( \widetilde{\omega}) \\
    g(X) &= \frac{X}{[1 + X^2]^{5/4}}. \label{eq:gki_gx}
\end{align}
The scaling relations in Eqs. (\ref{eq:re_fxc_gki})--(\ref{eq:gki_gx}) greatly simplify the numerical evaluation of the kernel, although they are believed to be exact only within the GKI frequency interpolation.
The dimensionless function $h(X)$ enforces these limits
\begin{eqnarray}
  \lim_{X \to 0} h(X) &\to& \frac{1}{\gamma} \\
  \lim_{\omega \to \infty} \re \fxc(0,\omega) &\to& \fxc(0,\infty) + \frac{c}{\omega^{3/2}}
\end{eqnarray}
while modeling the finite frequency dependence of $\re \fxc(0,\omega)$ through the Kramers-Kronig principal value integral. As noted in the Introduction, repeated evaluation of $\re \fxc(0,\omega)$ through the Kramers-Kronig integral is computationally expensive. Therefore, an accurate model of the Kramers-Kronig-derived frequency dependence through $h$ is an essential component of an analytic and numerically efficient $\fxc(0,\omega)$. Figure 4 of Ref. \cite{ruzsinszky2020} shows that $h$ adequately models this frequency dependence, however $h$ can be improved. We propose a simple modification to the MCP07 $h(X)$ function
\begin{equation}
  h(X) = \frac{1}{\gamma}\frac{1 - c_1 X^2}{[1 + c_2 X^2 + c_3 X^4 + c_4 X^6 + (c_1/\gamma)^{16/7} X^8]^{7/16}} \label{eq:new_hx},
\end{equation}
where the parameters
\begin{align}
    (c_1,c_2,c_3,c_4) = (& 0.174724,3.224459,2.221196, \nonumber \\
    & 1.891998)
\end{align}
were determined by directly fitting to numeric Kramers-Kronig results. Note that $h$ is an even function of real-valued frequency.
(An exact expression for $h$ is given in Eq. 4.84 of Ref. \cite{primer_tddft}, however this expression involves nonstandard special functions.)

We also need to analytically continue the GKI kernel to imaginary frequencies. As this case is useful for the evaluation of the correlation energy, the analytic continuation to purely imaginary frequencies can be accurately represented by
\begin{align}
  & f\sxc(0,i u) \approx - c [b(n)]^{3/4} j(\widetilde{\omega}) + f(0,\infty) \\
  & j(y) = \frac{1}{\gamma}\frac{1 - k_1 y + k_2 y^2}{[1 + k_3 y^2 + k_4 y^4 + k_5 y^6 + (k_2/\gamma)^{16/7} y^8]^{7/16}}
\end{align}
with the $k_i$,
\begin{align}
    (k_1,k_2,k_3,k_4,k_5) = (& 1.219946,0.973063,0.42106,\nonumber \\
      & 1.301184,1.007578),
\end{align}
determined by a non-linear least-squares fit to an $\rs$-independent form, followed by a grid search to refine the parameters. $u\geq 0$ is purely real.

In this work, we will use the Perdew-Wang parametrization \cite{perdew1992} of the correlation energy per electron in jellium, as this yields an improved, smoother fit to quantum Monte-Carlo data \cite{ceperley1980} than does the Perdew-Zunger parametrization \cite{perdew1981} used for $\fxc(0,0)$ in the MCP07 kernel. Reference \cite{perdew2021} also made it clear that the MCP07 kernel does not adequately reproduce the correlation energies per electron in jellium at intermediate densities ($10 < \rs < 100$). The correlation energy per particle is given by the multi-dimensional integral \cite{langreth1977}
\begin{equation}
  \varepsilon\suc = \frac{1}{2} \int \frac{d^3 q}{(2\pi)^3} \int_0^1 \frac{d\lambda}{\lambda} \int_0^{\infty} d\omega \frac{4\pi \lambda }{q^2}[ S_{\lambda}(\qv,\omega) - S_0(\qv,\omega) ], \label{eq:eps_c_acdft}
\end{equation}
where $f_{\mathrm{xc},\lambda}(\qv,\omega,\rs) = \lambda^{-1}\fxc(\lambda^{-1} \qv, \lambda^{-2} \omega, \lambda \rs)$ \cite{lein2000} and $S_{\lambda}$ is evaluated using the coupling-constant $\lambda$-scaled $\fxc$. Note that $S_0(\qv,\omega) = -\im \xks/(\pi n)$. We adopt a similar integration scheme as Ref. \cite{perdew2021} to evaluate correlation energies per particle, but use a grid with a fixed number of points chosen to recover the RPA values reported there.

The ``screening'' wavevector $k$ in Eq. (\ref{eq:mcp07}) for the dynamic MCP07 kernel was chosen to be identical to the wavevector appearing in the static part of the MCP07 kernel. That choice was made consistent with an Occam's Razor-style construction principle: free parameters should be avoided when possible.

\begin{figure*}
  \centering
  \includegraphics[width=0.8\textwidth]{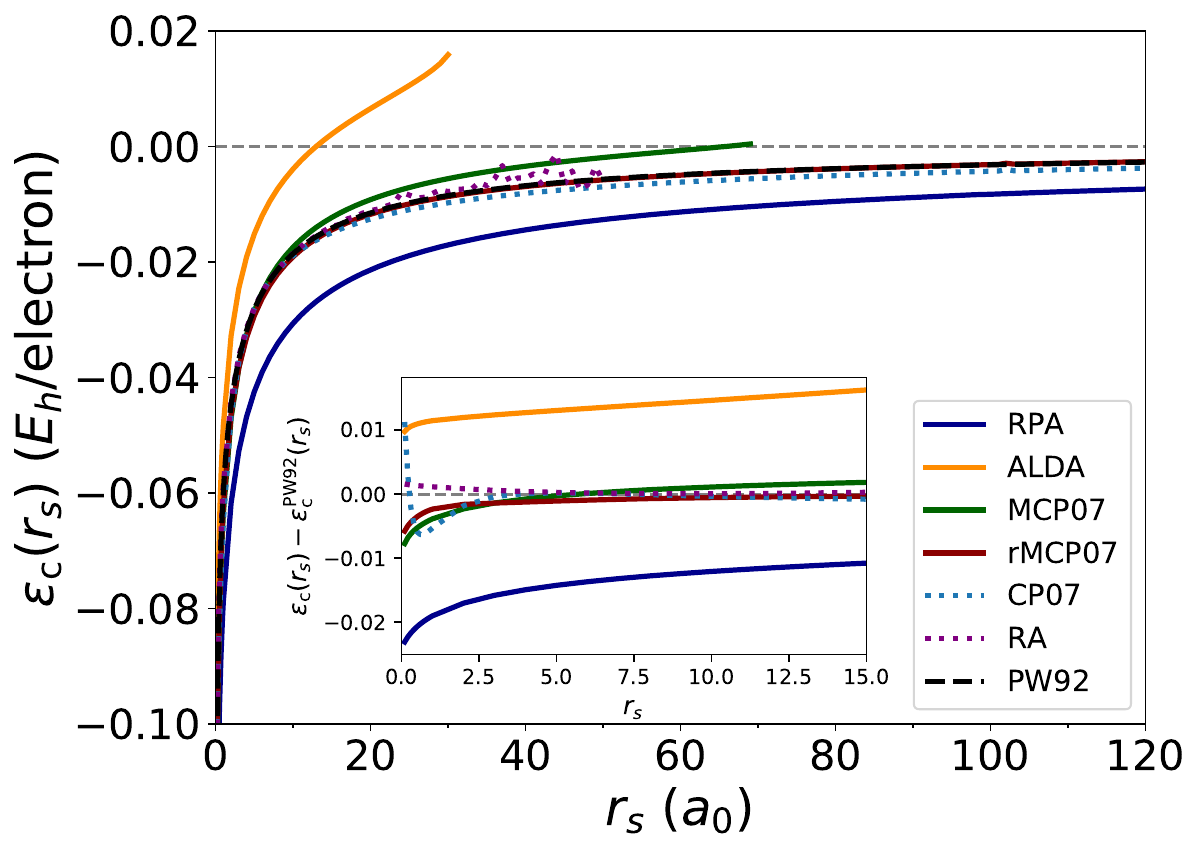}
  \caption{Demonstrating the higher accuracy of the rMCP07 kernel in predicting jellium correlation energies per electron (in units of hartree, $E_h$, per electron; note that $1 E_h \approx 27.211$ eV) at a range of density parameters $\rs$ (in units of bohr radii $a_0\approx 0.529$ \AA{}).
  Also depicted are the values computed with the Constantin-Pitarke (CP07) \cite{constantin2007} kernel and Richardson-Ashcroft (RA) \cite{richardson1994,lein2000} local field factor (see Eq. \ref{eq:lff_symm}).
  The inset plots the range $0 < \rs \leq 15$.
  PW92 \cite{perdew1992} (black, dashed) is essentially exact. For the values plotted here in the range $0.1 \leq \rs \leq 10$, see Table \ref{tab:ueg_eps_c}.
  Unlike CP07 and rMCP07, MCP07 is not fitted to the correlation energy.
  \label{fig:ueg_eps_c}}
\end{figure*}

Consider the revision
\begin{align}
  \fxc^{\text{rMCP07}}(q,\omega) &= \left\{1 + e^{-(q/\kt)^2} \left[ \frac{\fxc(0,\Omega)}{\fxc(0,0)} - 1\right] \right\} \nonumber \\
  & \times \fxc^{\text{MCP07}}(q,0) \label{eq:fxc_tc} \\
  \Omega &= p(\rs,q)\omega \\
  \kt &= \kf\frac{A + B \kf^{3/2}}{1 + \kf^2}, \label{eq:kscr} \\
  p(\rs,q) &= \left(\frac{\rs}{C}\right)^2 + \left[1 - \left(\frac{\rs}{C}\right)^2\right]\exp[-D(q/\kt)^2]. \label{eq:pscl}
\end{align}
The density dependence of $\kt$ will be discussed below. $p(\rs,q)$ is designed to tend to one as $q\to 0$, but to become much greater than one when $\rs \to \infty$ with $q > 0$. Moreover, the product $\rs^2\omega$ has no $\lambda$-dependence under the coupling-constant integration of Eq. (\ref{eq:eps_c_acdft}). Here
\begin{align}
    (A,B,C,D) = (&3.846991, 0.471351, 4.346063, \nonumber \\
      & 0.881313)
\end{align}
were determined by minimizing the unweighted sum
\begin{equation}
  \sigma = \sum_{\rs} |\varepsilon\suc^{\mathrm{rMCP07}}(\rs) - \varepsilon\suc^{\mathrm{PW92}}(\rs)|.
\end{equation}
For the fit, 20 values of $\rs$ in the range $1\leq \rs \leq 100$ bohr were used to determine $A$, $B$, $C$, and $D$. Over-fitting is avoided by using a large number of $\rs$ values and a fixed integration grid, where numeric convergence is not guaranteed to identical precision for each $\rs$. Figure S10 of Ref. \cite{perdew2021} shows that $\varepsilon\suc^{\mathrm{MCP07}}$ is least accurate at intermediate $\rs$, motivating the factor of $\rs^2$ in Eq. (\ref{eq:pscl}).
The accuracy of the rMCP07 kernel at intermediate densities is greatly improved, as seen in Fig. \ref{fig:ueg_eps_c}.
The rMCP07 kernel also represents an accurate extrapolation to $\rs > 100$ and $\rs < 1$.
From Fig. \ref{fig:ueg_eps_c}, we also see that rMCP07 improves upon the CP07 kernel at low densities, where CP07 predicts too-negative correlation energies, and at higher densities, where CP07's behavior is erratic.
At highest densities, the Richardson-Ashcroft local field factor \cite{richardson1994} (with corrections from Ref. \cite{lein2000}) is most accurate, but its accuracy degrades substantially as $\rs$ increases.

At low densities, exchange and correlation have the same length scale, the Fermi wavelength $2\pi/\kf$. Accordingly, at low densities, $\kt \propto \kf$. At high densities, the appropriate length scale for correlation is the inverse of the Thomas-Fermi wavevector, $k_{\text{TF}} = \sqrt{4\kf/\pi}$. Thus, $\kt \propto k_{\text{TF}}$ at high densities. These effects are built into Eq. (\ref{eq:kscr}).

There is existing precedence for scaling the frequency-dependent part of the kernel by a function of $q$, as we have by introducing $\Omega(\rs,q,\omega)$. Dabrowski \cite{dabrowski1986} sought to extend the long-wavelength Gross-Kohn kernel \cite{gross1985} to nonzero $q$ by enforcing zero and infinite \cite{niklasson1974} frequency limits on the spin-symmetric local field factor \cite{giuliani2005}
\begin{equation}
  G_+(q,\omega)=\frac{1}{2}[G_{\uparrow\uparrow}(q,\omega)+G_{\uparrow\downarrow}(q,\omega)] =-\frac{q^2}{4\pi}\fxc(q,\omega). \label{eq:lff_symm}
\end{equation}
The Dabrowski kernel is limited in that it uses older expressions for the static local field factors \cite{vashishta1972,pathak1973,utsumi1980} which have no closed form, and predated the work of Iwamoto and Gross \cite{iwamoto1987}, which corrected the Gross-Kohn expression to enforce the third frequency-moment sum rule.

It should also be noted that the spin-antisymmetric local field factor $G_-(q,\omega) = [G_{\uparrow\uparrow}(q,\omega)-G_{\uparrow\downarrow}(q,\omega)]/2$ is needed to describe the spin-spin response function
\begin{equation}
  \frac{\xks}{1 - 4\pi/\qv^2[1-G_-(q,\omega)]\xks}.
\end{equation}
At present, we lack sufficient information to determine a first-principles, spin-polarized $\fxc$ from the uniform electron gas. Works like those of Richardson and Ashcroft \cite{richardson1994} are therefore useful in understanding the spin-spin response, which is needed to describe two-electron interactions \cite{kukkonen1979}, such as those that spur formation of Cooper pairs. It is important to note that the full correlation energy is still included in $\fxc(q,\omega)$, even if it is not decomposed into same- and opposite-spin components. This is in stark contrast to some approximate expressions for $G_+$ which assume $G_{\uparrow\downarrow}\approx 0$, thereby neglecting at least opposite-spin correlation interactions. A spin decomposition of the ALDA is given in Ref. \cite{gori2004}.

Our kernel retains the broad features of these earlier works. It may well be possible to enforce known limits on $G(q,\omega)$, however all existing work is $\rs$-dependent, primarily in a metallic range $1 \lesssim \rs \lesssim 10$. Real solids have regions of significant density depletion (e.g., vacancies and voids in semiconductors). By constraining the model kernel to recover accurate jellium energetics at a wide range of densities, we hope to better describe real systems.

\begin{ruledtabular}
  \begin{table}[t]
      \centering
      \begin{tabular}{rrrrrr}
  $\rs$ & $\varepsilon_{\mathrm{c}}$ PW92 & RPA & ALDA & MCP07 & rMCP07 \\ \hline 0.1 & -0.1209 & -0.1440 & -0.1111 & -0.1286 & -0.1267 \\
  0.2 & -0.1011 & -0.1234 & -0.0908 & -0.1079 & -0.1061 \\
  0.3 & -0.0900 & -0.1117 & -0.0794 & -0.0962 & -0.0944 \\
  0.4 & -0.0824 & -0.1035 & -0.0716 & -0.0881 & -0.0863 \\
  0.5 & -0.0766 & -0.0973 & -0.0657 & -0.0819 & -0.0802 \\
  0.6 & -0.0720 & -0.0923 & -0.0609 & -0.0770 & -0.0753 \\
  0.7 & -0.0682 & -0.0882 & -0.0570 & -0.0729 & -0.0712 \\
  0.8 & -0.0650 & -0.0846 & -0.0537 & -0.0694 & -0.0677 \\
  0.9 & -0.0622 & -0.0815 & -0.0508 & -0.0663 & -0.0647 \\
  1 & -0.0598 & -0.0788 & -0.0483 & -0.0636 & -0.0621 \\
  2 & -0.0448 & -0.0618 & -0.0328 & -0.0471 & -0.0464 \\
  3 & -0.0369 & -0.0528 & -0.0246 & -0.0383 & -0.0383 \\
  4 & -0.0319 & -0.0468 & -0.0191 & -0.0326 & -0.0331 \\
  5 & -0.0282 & -0.0425 & -0.0152 & -0.0285 & -0.0293 \\
  6 & -0.0254 & -0.0391 & -0.0120 & -0.0253 & -0.0264 \\
  7 & -0.0232 & -0.0364 & -0.0095 & -0.0228 & -0.0240 \\
  8 & -0.0214 & -0.0342 & -0.0074 & -0.0207 & -0.0221 \\
  9 & -0.0199 & -0.0323 & -0.0055 & -0.0190 & -0.0205 \\
  10 & -0.0186 & -0.0307 & -0.0039 & -0.0175 & -0.0191 \\
      \end{tabular}
      \caption{Jellium correlation energies per particle $\varepsilon\suc$, in hartree/electron, for a variety of XC kernels and reference PW92 \cite{perdew1992} values. For a plot of $\varepsilon\suc$ on the range $0.1 \leq \rs \leq 120$, see Fig. \ref{fig:ueg_eps_c}. The values of $\varepsilon\suc$ were determined using a denser integration grid than was used to fit the rMCP07 parameters.}
      \label{tab:ueg_eps_c}
  \end{table}
\end{ruledtabular}

A similar approach was taken by Panholzer {\it et al.} \cite{panholzer2018}, who directly tabulated highly accurate expressions for $\fxc(q,\omega)$ in jellium at a range of densities $0.8 \leq \rs \leq 8$, frequencies and wavevectors, as well as a prescription for using it in real systems (a ``connector''). Many-body theory approaches can also be used to tabulate the dielectric function of jellium, as was done in Ref. \cite{chen2019} for the static response. Our approach may yield greater generality.

\begin{figure}
  \centering
  \includegraphics[width=\columnwidth]{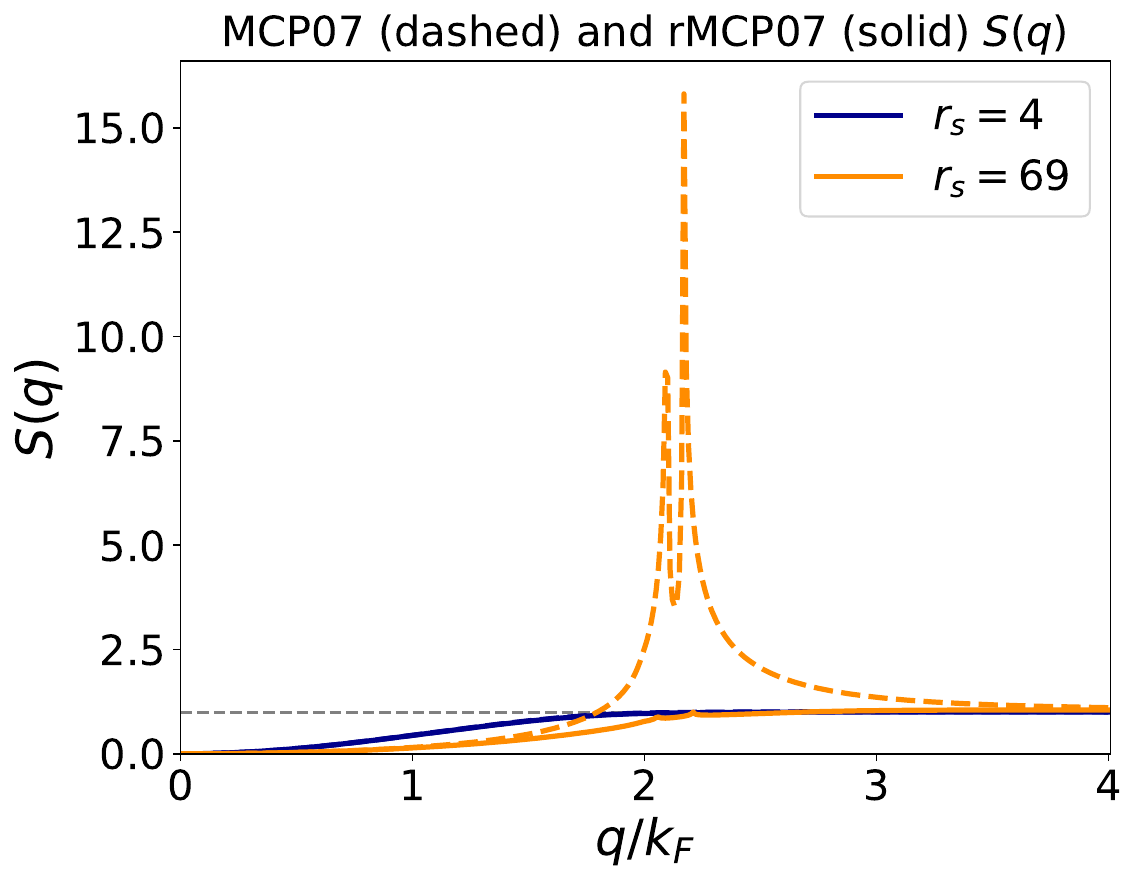}
  \caption{Comparison of the static structure factors $S(q)$ for the MCP07 (dashed) and rMCP07 (solid) kernels at a higher, $\rs = 4$ (blue), and much lower, $\rs = 69$ (orange), density. The rMCP07 kernel almost completely eliminates the unphysically large peak structure seen in the MCP07 kernel at lower densities. For a plot of the rMCP07 static structure factor alone, see Fig. \ref{fig:s_q_new}. \label{fig:s_q_comp} }
\end{figure}

\begin{figure}
  \centering
  \includegraphics[width=\columnwidth]{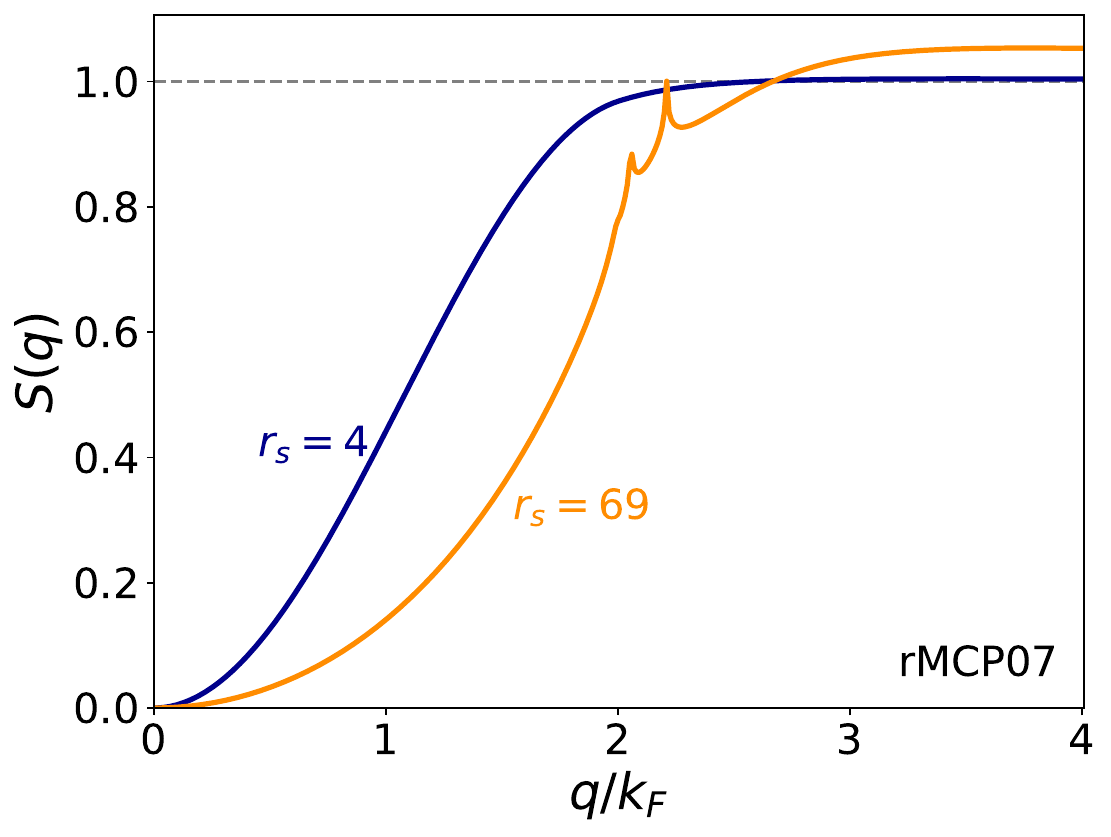}
  \caption{The static structure factor $S(q)$ of the rMCP07 kernel. \label{fig:s_q_new}}
\end{figure}

These modifications also soften the peak structure seen in $S(q)$ of Eq. (\ref{eq:sq_def}) for $\rs = 69$. Figures \ref{fig:s_q_comp} and \ref{fig:s_q_new} show clearly that the large MCP07 peak in the $\rs=69$ curve is reduced substantially, while the $\rs = 4$ curve is essentially unchanged. It is difficult to determine what $S(q)$ should look like at all densities. A parameterization of the jellium $S(q)$ from QMC data for $r_s \leq 10$ \cite{gori2000} suggests a monotonic increasing $S(q)$ at most densities.
At intermediate densities, this parameterization represents an extrapolation of unknown accuracy; previously unpublished QMC data \cite{ortiz1999} at lower densities suggests that $S(q)$ is nonmonotonic, as shown in Fig. \ref{fig:qmc_dat} of Appendix B.

\section{Characterizing the \MakeLowercase{r}MCP07 kernel}

\subsection{Static charge density wave in jellium}

\begin{figure}
    \centering
    \includegraphics[width=\columnwidth]{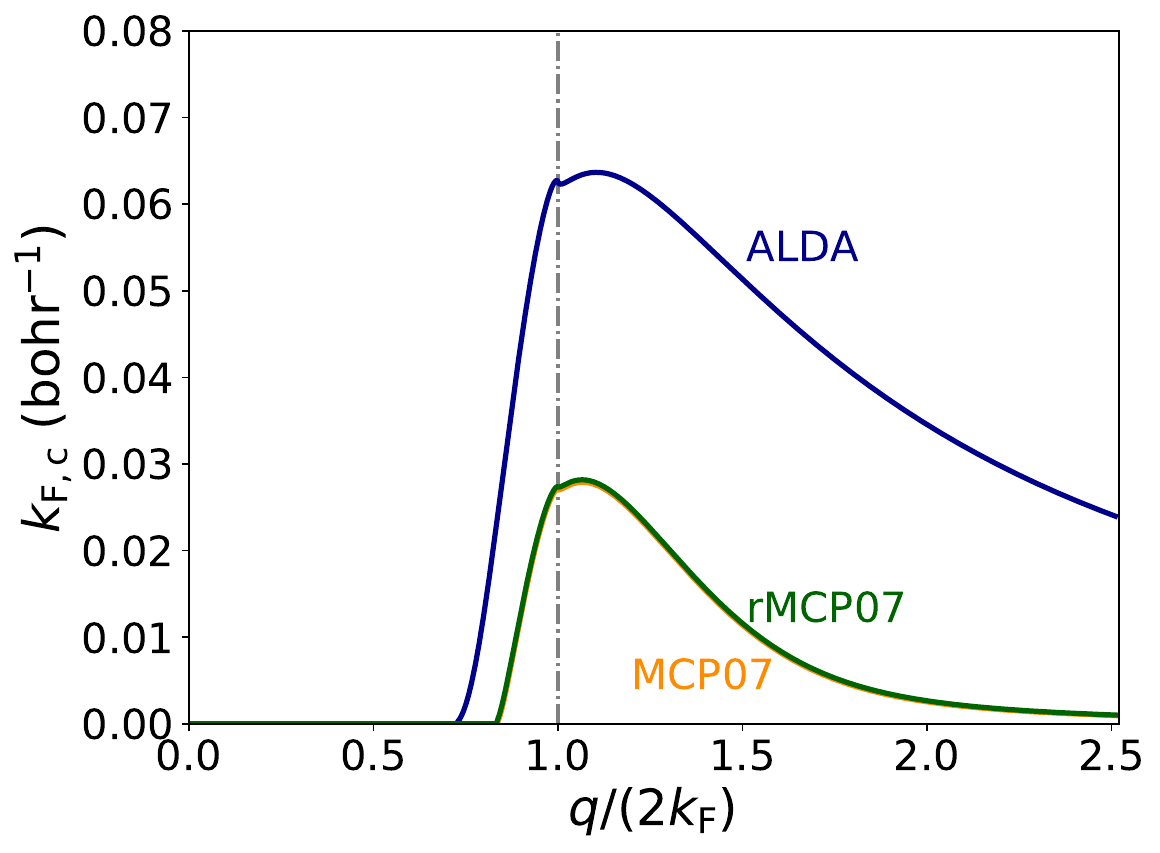}
    \caption{Plot of the critical Fermi wavevector $k_{\mathrm{F,c}}$, or equivalently, critical Wigner-Seitz radius $r_{\mathrm{s,c}}$, such that the static dielectric function of Eq. (\ref{eq:df_static}) vanishes in jellium, signaling possible onset of a static charge density wave. For the RPA $k_{\mathrm{F,c}}=0$ at seemingly all wavevectors considered here.}
    \label{fig:kfc}
\end{figure}

Here we will discuss the appearance of a static charge-density wave in jellium at low density. A first-order phase transition often occurs close to a singularity in a linear response function, in our case $\chi(\qv,\omega)$ of Eq. (\ref{eq:chi_recip}). Let $k_{\mathrm{F,c}}$ be the critical Fermi wavevector [and $r_{\mathrm{s,c}}=(9\pi/4)^{1/3}/k_{\mathrm{F,c}}$] such that the static dielectric function
\begin{equation}
    \widetilde{\epsilon}[n](q,0) = 1 - \left[\frac{4\pi}{q^2} + \fxc[n](q,0) \right]\chi_0[n](q,0) \label{eq:df_static}
\end{equation}
vanishes. The results of this calculation, comparable to Fig. 2 of Ref. \cite{ruzsinszky2020}, are shown in Fig. \ref{fig:kfc}. As reported there, we find that $r_{\mathrm{s,c}}\approx 30$ for the ALDA, and $r_{\mathrm{s,c}}\approx 69$ for MCP07; for rMCP07, $r_{\mathrm{s,c}}\approx 68$, exceedingly similar to MCP07. It should be noted that MCP07 and rMCP07 do not have exactly the same static limits because of the different parameterizations of the ALDA used.

\subsection{Sum rules}

An important set of constraints on the spectral function are frequency-moment sum rules of the form
\begin{equation}
  \Sigma_M(q) \equiv \int_0^{\infty} \omega^M S(q,\omega)d\omega,
\end{equation}
where $\Sigma_M $ is ostensibly known. For example, the ``$f$-sum'' rule (see Eq. 3.141 of Ref. \cite{giuliani2005}) states that the first frequency moment, in jellium
\begin{equation}
  \Sigma_1(q) = \frac{q^2}{2},
\end{equation}
which was already well-satisfied by MCP07 \cite{perdew2021}. Reference \cite{perdew2021} demonstrated that MCP07 struggled with the third frequency-moment sum rule (see Eq. 3.142 of Ref. \cite{giuliani2005})
\begin{align}
  &\Sigma_3(q) = \frac{q^2}{2}\left\{ \frac{q^4}{4} + 4\pi n + 2 q^2 \left(t_0 + t\suc \right) \right. \nonumber \\
  & \left. + \frac{1}{\pi} \int_0^{\infty} dk \int_{-1}^1 du ~k^2 u^2 [ S(\sqrt{q^2 + k^2 - 2k q u})-S(k)] \right\} \label{eq:m3_sr}
\end{align}
in jellium at low densities. In Eq. (\ref{eq:m3_sr}), $t_0=\frac{3}{10} \kf^2$ is the non-interacting kinetic energy per electron in jellium, and $t\suc$ is the interacting kinetic energy per electron.
$t\suc$ can be computed from the virial theorem \cite{levy1985}
\begin{equation}
  t\suc = -4 \varepsilon\suc(\rs,0) + 3v\suc(\rs,0), \label{eq:tc_virial}
\end{equation}
where $\varepsilon_c(\rs,\zeta)$ is the correlation energy per electron of jellium, $v\suc=\partial(n \varepsilon\suc)/\partial n$ is the corresponding (g.s.) correlation potential, and $\zeta = (n_{\uparrow}-n_{\downarrow})/n$ is the relative spin-polarization, which we take to be zero.
To evaluate $t\suc$, we use the parameterization of $\varepsilon\suc(\rs,\zeta)$ given by Ref. \cite{perdew1992}.

\begin{figure}[!ht]
  \centering
  \includegraphics[width=\columnwidth]{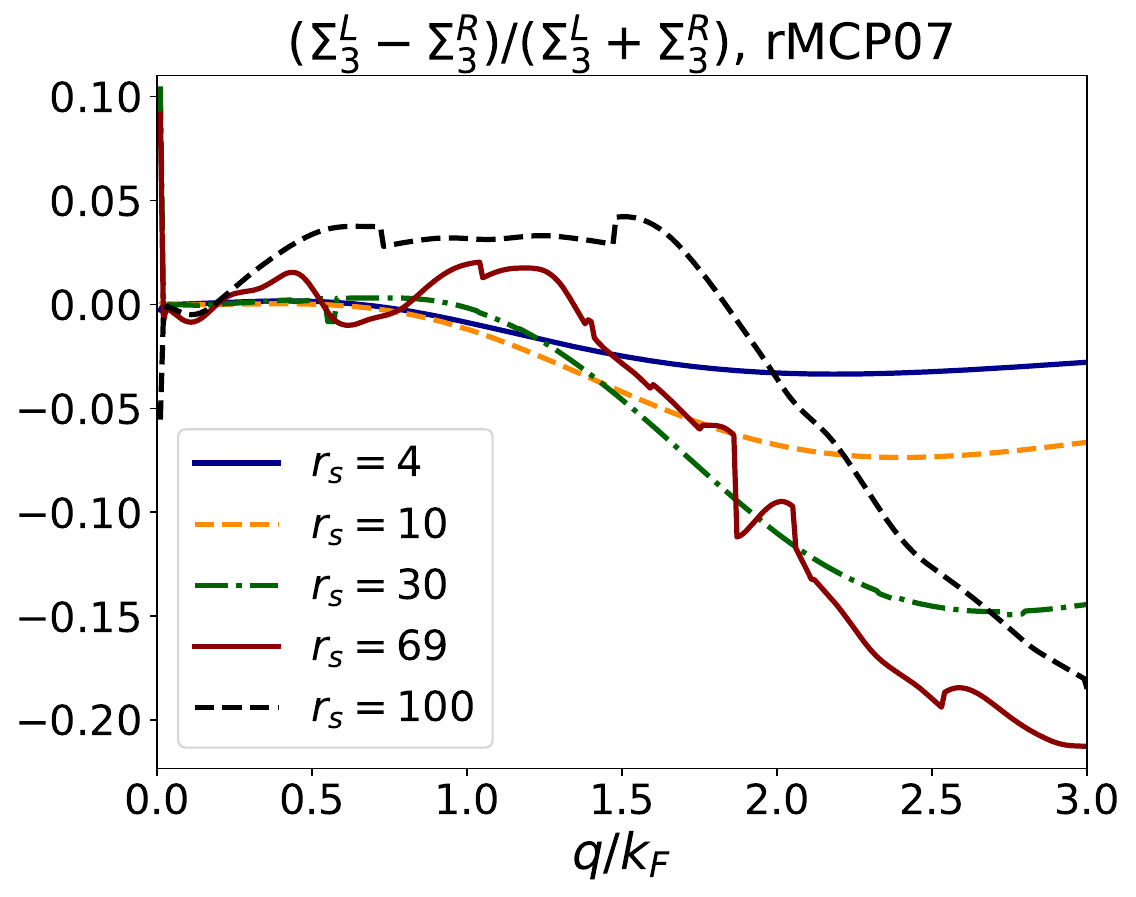}
  \caption{Relative differences in the third-frequency moment sum rule of Eq. (\ref{eq:m3_sr}) for rMCP07. $\Sigma_3^L$ represents the left-hand side of Eq. (\ref{eq:m3_sr}) [$\int_0^{\infty} \omega^3 S(q,\omega) d\omega$], and $\Sigma_3^R$ the right-hand side of Eq. (\ref{eq:m3_sr}). The third moment sum rule is satisfied nearly exactly by rMCP07 at a wide range of densities of jellium. \label{fig:m3_sr_new}}
\end{figure}

The rMCP07 kernel satisfies the third moment sum rule nearly exactly at a range of densities, as shown in Fig. \ref{fig:m3_sr_new}. This figure was generated in much the same way as Fig. S9 of Ref. \cite{perdew2021}, however the integration cutoff was set to $k_{\mathrm{c}}=14\kf$, much larger than the cutoff used there ($\sim 4\kf$). Moreover, a careful extrapolation to $k > k_{\mathrm{c}}$ was made in this work.

\begin{figure}
  \centering
  \includegraphics[width=\columnwidth]{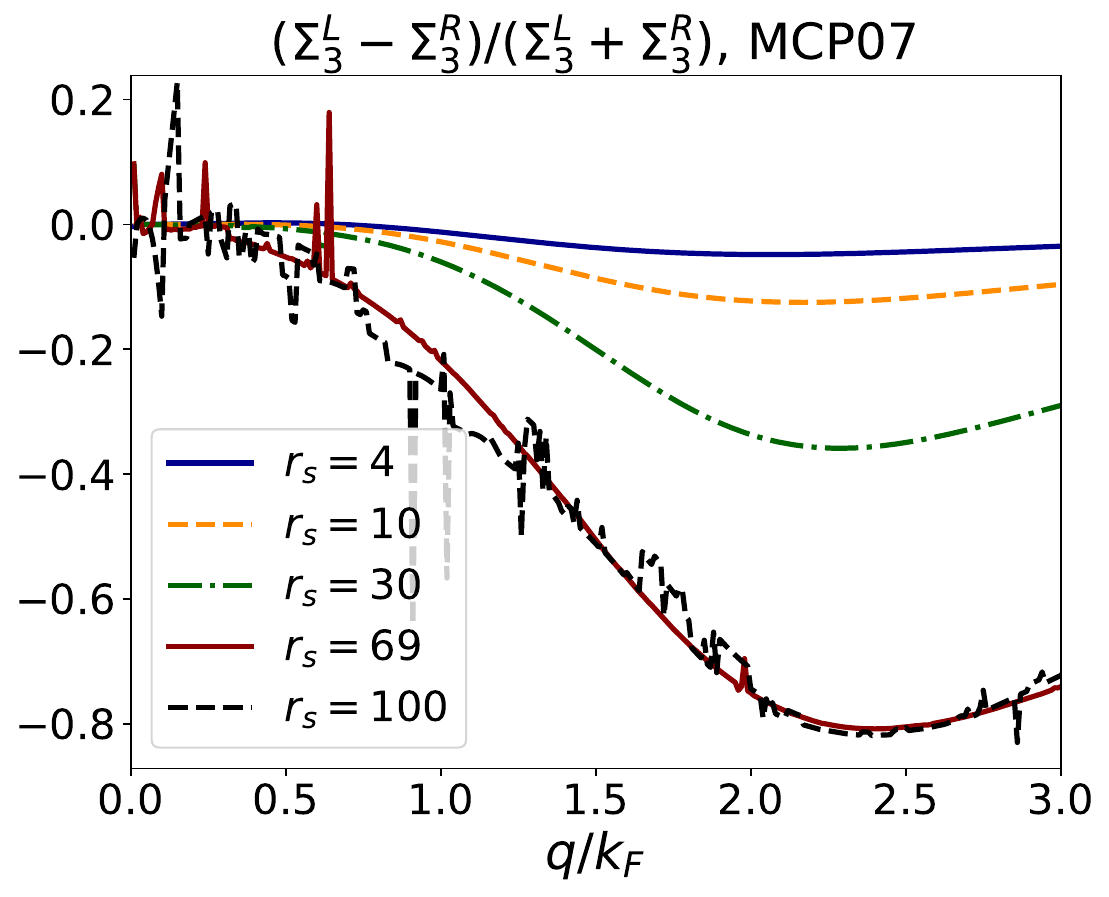}
  \caption{Relative differences in the third-frequency moment sum rule of Eq. (\ref{eq:m3_sr}) for MCP07. $\Sigma_3^L$ represents the left-hand side of Eq. (\ref{eq:m3_sr}) [$\int_0^{\infty} \omega^3 S(q,\omega) d\omega$], and $\Sigma_3^R$ the right-hand side of Eq. (\ref{eq:m3_sr}). The third moment sum rule is satisfied only approximately in MCP07 at intermediate to low density jellium. These results use a higher integration cutoff $k_{\mathrm{c}}=14\kf$ for $\rs \geq 10$ jellium. \label{fig:m3_sr_MCP07}}
\end{figure}

For comparison, Fig. \ref{fig:m3_sr_MCP07} shows the relative differences in the left- and right-hand sides of Eq. (\ref{eq:m3_sr}) computed with MCP07 using the higher cutoff. (Since neither the left nor the right sides of Eq. (\ref{eq:m3_sr}) are known exactly, the standard relative error cannot be calculated here.) Note that, for both the MCP07 kernel and the rMCP07 kernel, increasing the cutoff to $30\kf$ introduces large numeric instabilities in the integration. The maximum errors made by both kernels are tabulated in Table \ref{tab:m3_sr_amax}.

\begin{ruledtabular}
  \begin{table}
    \centering
      \begin{tabular}{r|rrrr}
        $\rs$ & MURD MCP07 & $q_{\text{MURD}}/\kf$ & MURD rMCP07 & $q_{\text{MURD}}/\kf$ \\ \hline
        4 & 0.048 & 2.06 & 0.034 & 2.19 \\
        10 & 0.125 & 2.16 & 0.074 & 2.40 \\
        30 & 0.358 & 2.29 & 0.149 & 2.74 \\
        69 & 0.808 & 2.42 & 0.213 & 3.00 \\
        100 & 0.830 & 2.86 & 0.185 & 3.00 \\
      \end{tabular}
      \caption{Comparison of the maximum unsigned relative differences (MURD) for MCP07 and rMCP07 in the third moment sum rule calculation, and the corresponding value of $q_{\text{MURD}}/\kf$ where the maximum occurs. As shown in Figs. \ref{fig:m3_sr_new} and \ref{fig:m3_sr_MCP07}, the relative difference is defined as  the difference between the left and right hand sides of Eq. (\ref{eq:m3_sr}), divided by their sum. \label{tab:m3_sr_amax}}
  \end{table}
\end{ruledtabular}

\subsection{Dressed interaction}

\begin{figure}[!ht]
  \includegraphics[width=\columnwidth]{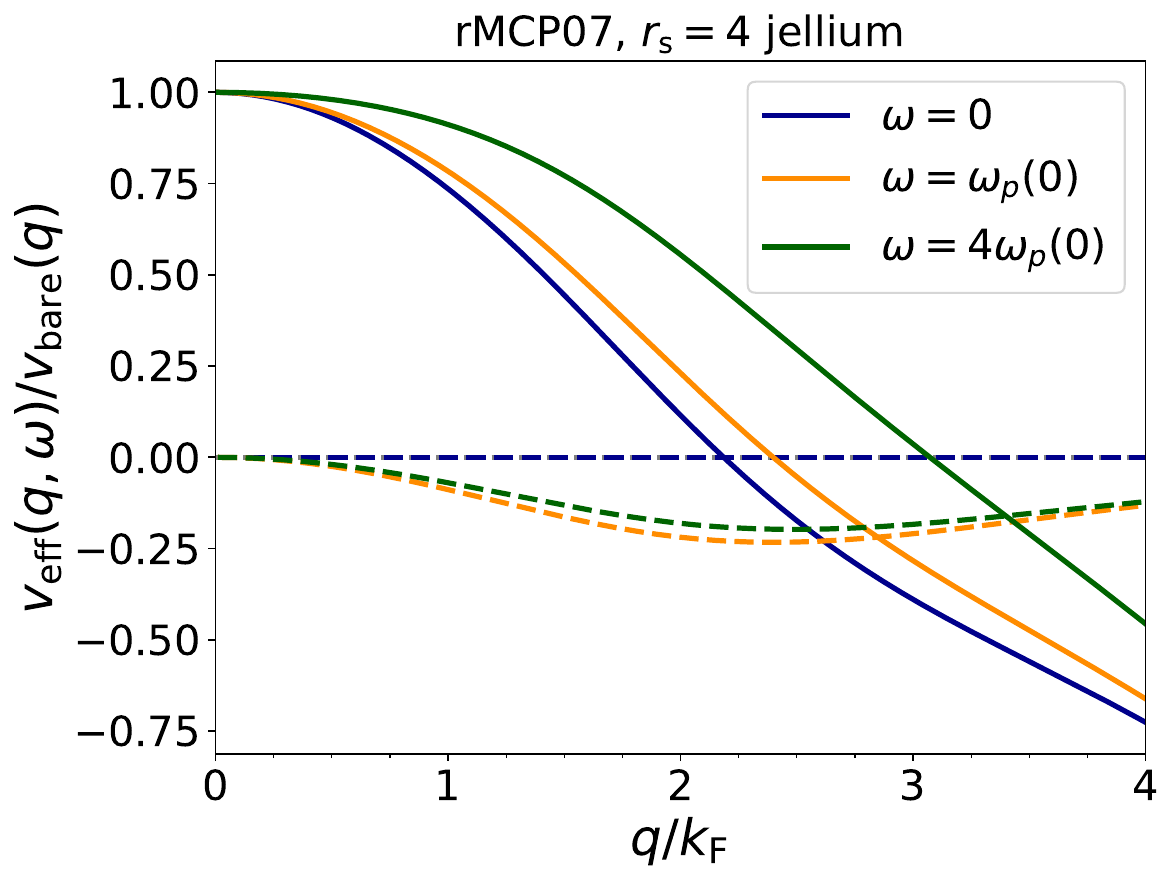}
  \caption{Real (solid) and imaginary (dashed) parts of the scaled effective potential $\veff/\vb$ for $\rs=4$ bulk jellium with the rMCP07 kernel. The crossings are
  $\re \veff(2.185\kf,0) = 0$, $\re \veff(2.398\kf,\omega_p(0)) = 0$, and $\re \veff(3.072\kf,4\omega_p(0)) = 0$.
  \label{fig:rmcp07_veff_rs_4}}
\end{figure}

\begin{figure}[!ht]
  \includegraphics[width=\columnwidth]{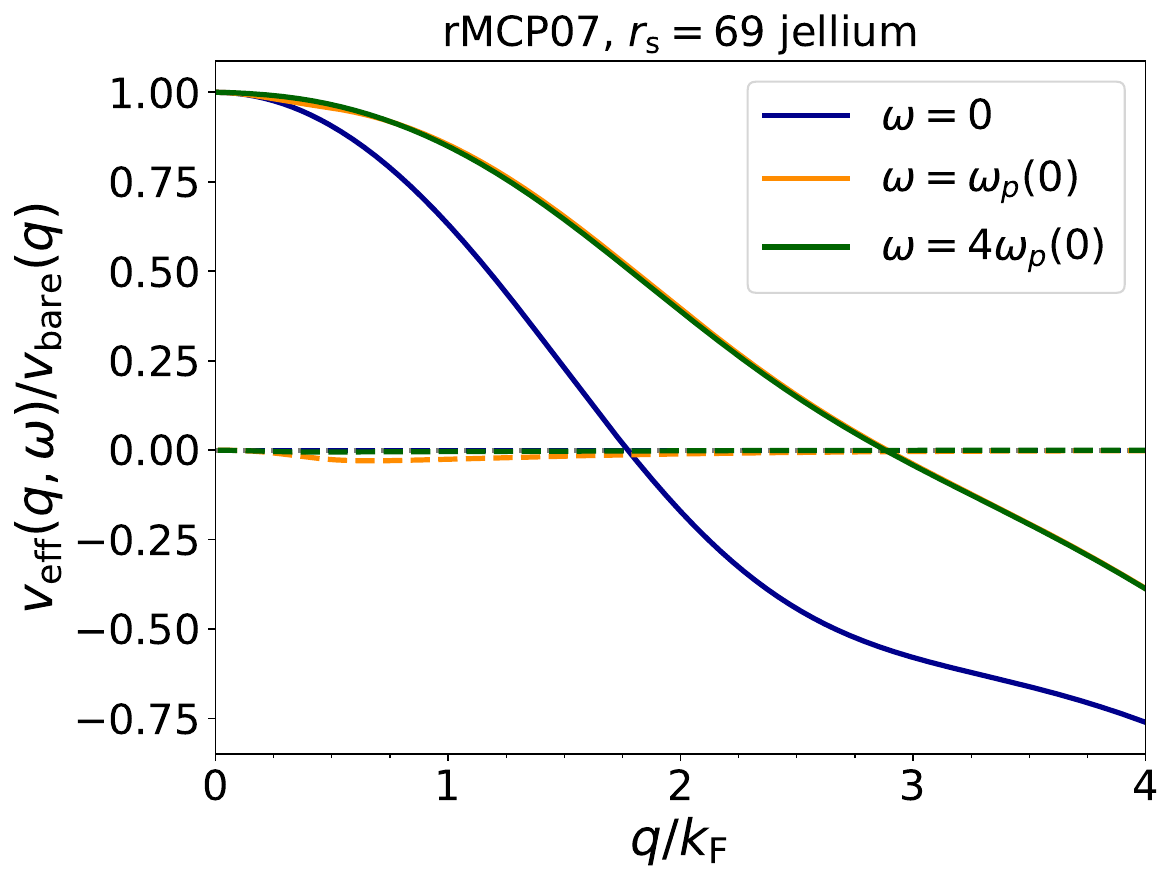}
  \caption{Real (solid) and imaginary (dashed) parts of the scaled effective potential $\veff/\vb$ for $\rs=69$ bulk jellium with the rMCP07 kernel. The crossings are
  $\re \veff(1.773\kf,0) = 0$, $\re \veff(2.889\kf,\omega_p(0)) = 0$, and $\re \veff(2.879\kf,4\omega_p(0)) = 0$. \label{fig:rmcp07_veff_rs_69}}
\end{figure}

Within density response theory, the dressed interaction (the effective
electron-electron interaction that makes the random
phase approximation exact),
\begin{equation}
  \veff(q,\omega) = \vb(q) + \fxc(q,\omega),
\end{equation}
where the bare interaction is $\vb(q) = 4\pi/q^2$, is of central importance, as shown by Eq. (\ref{eq:chi_recip}). As $q$ grows large, it is possible for $\veff$ to become negative; similarly, the dielectric function
\begin{equation}
  \widetilde{\epsilon}(q,\omega) = 1 - \veff(q,\omega)\chi_0(q,\omega) \label{eq:df}
\end{equation}
may become negative, as seen in Figs. \ref{fig:TC_epst_rs_4} and \ref{fig:TC_epst_rs_69} of the Appendix. The dressed interactions are plotted for the rMCP07 kernel at $\rs=4$ and 69 in Figs. \ref{fig:rmcp07_veff_rs_4} and  \ref{fig:rmcp07_veff_rs_69} respectively. At metallic densities and at intermediate densities, the effective potential becomes attractive only for $q \gtrsim \kf$.

The scaled frequency $\Omega$ entering rMCP07 is greater than the frequency $\omega$ for densities $\rs > C$. Thus, at lower densities, the rMCP07 kernel more rapidly approaches the infinite frequency limit than does MCP07. These differences are discernible in the dressed interaction at metallic densities. Moreover, as $\rs$ increases, the differences become more pronounced, as $\Omega$ grows with $\rs^2$ for $q \gtrsim \kt$. For example, at $\rs=69$, the rMCP07 dressed interaction has approached its infinite frequency limit for $\omega \approx \omega_p(0)$, whereas the MCP07 kernel tends closely to its static limit for $\omega = \omega_p(0)$.

There are numerous interpretations of a negative dressed interaction or negative dielectric function \cite{dolgov1981}, so we mention only a few here. These conditions imply that the screened interaction is attractive, which may underpin unconventional mechanisms of superconductivity. The Kohn-Luttinger \cite{kohn1965a} theory posits that Friedel oscillations (characteristic of jellium and simple metal surfaces) lead to regions of attractive dressed interactions, allowing for Cooper pairing without consideration of electron-phonon interactions. A first-principles description of superconductivity using a $\veff(q,\omega)$ derived from a well-constrained local field factor \cite{richardson1994} was developed by Richardson and Ashcroft \cite{richardson1997}. For a phenomenological review of attractive quasiparticle interactions, see Ref. \cite{monthoux2007}; for the relationship between the dielectric function and high-$T_{\mathrm{c}}$ superconductors, see Ref. \cite{dolgov1981}.

A collective mode corresponding to $\widetilde{\epsilon}(q)<0$, where $\widetilde{\epsilon}(q)$ is the static dielectric function, has been called a ``ghost plasmon'' \cite{takayanagi1997}, and it was found that this mode competes with the plasmon mode at intermediate densities, $\rs \approx 22$ \cite{takada2016}. Given that the mode emerges from poles of $\widetilde{\epsilon}(q,\omega)$ at conjugate imaginary frequencies \cite{takada2016}, this excitation is better labeled as an exciton. (The name ``ghost exciton'' is eye-catching, but badly obscures \emph{what} the collective mode represents. The original work \cite{takayanagi1997} found that the collective mode contributes dominantly to the first-frequency-moment sum rule, and destabilizes the system.)

Further work \cite{panholzer2018} showed that the exciton appeared in the ALDA static response, but not in the RPA response. Their work demonstrated that inclusion of two-particle, two-hole ($2p2h$) excitations in a Fermi hypernetted chain-correlated basis function calculation of bulk jellium indeed produces an excitonic mode at intermediate densities. Figure \ref{fig:ghost_exciton} of Appendix C shows that the MCP07 and rMCP07 kernels also miss this excitonic mode, but that the dynamic LDA of Qian and Vignale (QV) \cite{qian2002}, which satisfies a different static limit than the GKI dynamic LDA,  captures the excitonic mode. The QV kernel is discussed in Appendix C.

Consider instead the change in density $\delta n$ due to a weak external perturbation $\delta v_{\text{ext}}$. Linear response dictates that
\begin{equation}
    \delta n(q,\omega) = \xks \delta v_{\text{s}}(q,\omega) = \frac{\xks}{\widetilde{\epsilon}(q,\omega)}\delta v_{\text{ext}},
\end{equation}
where
\begin{equation}
    \delta v_{\text{s}}(q,\omega) = \delta v_{\text{ext}}(q,\omega)  + v_{\text{eff}}(q,\omega) \delta n(q,\omega)
\end{equation}
is the change in the Kohn-Sham potential due to the perturbation. $\delta v_{\text{s}}$ describes how the density screens $\delta v_{\text{ext}}$, and thus can be used to describe screening in real systems.

\section{Conclusions}

We have motivated, presented, and analyzed an exchange correlation kernel for use in TD-DFT and linear response calculations based on known exact constraints.
This form is tightly constrained to reproduce accurate jellium correlation energies at all densities, a feat at which many common exchange-correlation kernels (even MCP07) fail.
As jellium contains much of the essential physics of metals, we anticipate that the rMCP07 and MCP07 kernels will accurately describe properties of real metals.

Both MCP07 and rMCP07 approximate the kernel of the spin-unpolarized fluid phase of jellium.
At densities typical of valence electrons in metals, for which this phase is the ground-state, both kernels accurately model $\fxc$.
At much lower densities, the spin-unpolarized fluid, spin-polarized fluid, and Wigner crystal phases are all very close in energy.
The unpolarized fluid phase may only be meta-stable in this range, although a recent calculation shows it may be stable \cite{holzmann2020}.
At these lower densities, the MCP07 static structure factor deviates appreciably from that of the paramagnetic fluid phase.
rMCP07 is constructed as an improvement upon MCP07 at all densities, but especially at these lower densities where jellium displays strong correlation and symmetry breaking.
The wavevector- and frequency-dependent MCP07 \cite{perdew2021} and rMCP07 (Appendix \ref{sec:dens_fluc}) XC kernels correctly predict a drop in the spectral function toward zero frequency at the known wavevector of the incipient static charge density wave.

Our former interpretation \cite{perdew2021} of Anderson's explanation for symmetry breaking required that, at or near the critical density $n$ and wavevector $q$, 100\% of the spectral weight $S(q,\omega)$ should drop to zero frequency $\omega$, as in Appendix E. Our current and more defensible interpretation is that only a significant fraction of the spectral weight should drop to zero frequency.

The satisfaction of more exact constraints can sometimes worsen some predictions. While rMCP07 is clearly more accurate than MCP07 for the static structure factor, the correlation energy, and the third-moment sum rule at intermediate densities ($10< \rs< 100$), Figs. \ref{fig:TC_plas_disp}, \ref{fig:avg_dens_fluc}, and \ref{fig:stddev_dens_fluc} of the appendices suggest that MCP07 may be more correct than rMCP07 for the plasmon dispersion and in a qualitative sense for the spectral function $S(q,\omega)$ at $\rs =69$. Fig. \ref{fig:TC_epst_rs_69} shows that the rMCP07 dielectric function $\widetilde{\varepsilon}$ has an unexpected and possibly spurious zero (in its real part) at $\rs = 69$, $q \approx 2 \kf$, and $\omega = \omega_p(0)$, which MCP07 does not have. This would create not only a strong peak in $S(q,\omega)$ at $\omega = 0$, but also a strong peak at $\omega = \omega_p(0)$. Removing this second zero of $\widetilde{\varepsilon}$ might further improve the rMCP07 approximation to the exchange-correlation kernel of jellium.

The exchange-correlation kernel for a real material should of course reduce to the jellium kernel as the electron density becomes more uniform. Knowing this kernel for a real system would make exact the random phase approximation for the ground-state energy, and would also enable an accurate calculation of the optical absorption spectrum. The main difference arises in the $q\to 0$ limit, where the jellium kernel tends to a finite constant, while the kernel of a real system shows, at optical frequencies, an ultranonlocality or $q^{-2}$ divergence that is further discussed in Appendix D.
We find that in rMCP07 the coefficient of this divergence is extremely small for real simple metals.

A highly accurate approximation to the kernel for jellium is a step towards an accurate kernel for real metals, and ultimately for semiconductors and insulators. In the jellium limit, and in the density range $0 < \rs < 10$ important for real materials, the kernel $\fxc(n,q,\omega)$ is described well by MCP07 and even better by rMCP07, although both might be further improved by making a more realistic interpolation $\fxc(n,0,\omega)$ between the known high- and low-frequency limits (as discussed further in Appendix C). But this improvement would likely lose the closed-form analytic expression that makes the kernel potentially most useful.

The code used to fit the revised MCP07 kernel is made freely available at \cite{code_repo}. The data used to generate plots of the revised kernel are available in the ``published\_data'' directory of the code repository \cite{code_repo}.

\begin{acknowledgements}
  The work of ADK was supported by the Department of Energy, Basic Energy Sciences, under grant No. DE-SC0012575, and by Temple University. The work of NKN and AR was supported by the U.S. National Science Foundation (NSF) under Grant No. DMR-1553022. The work of JPP was supported by NSF Grant DMR-1939528, with a contribution from Chemical Theory, Modeling, and Computation, Division of Chemistry.
\end{acknowledgements}

\section*{Conflict of interest}

The authors declare that they have no financial and no non-financial conflicts of interest.

%

\appendix
\section{Plots of the rMCP07 dielectric function and related quantities}

\begin{figure*}[ht]
    \centering
    \includegraphics[width=\textwidth]{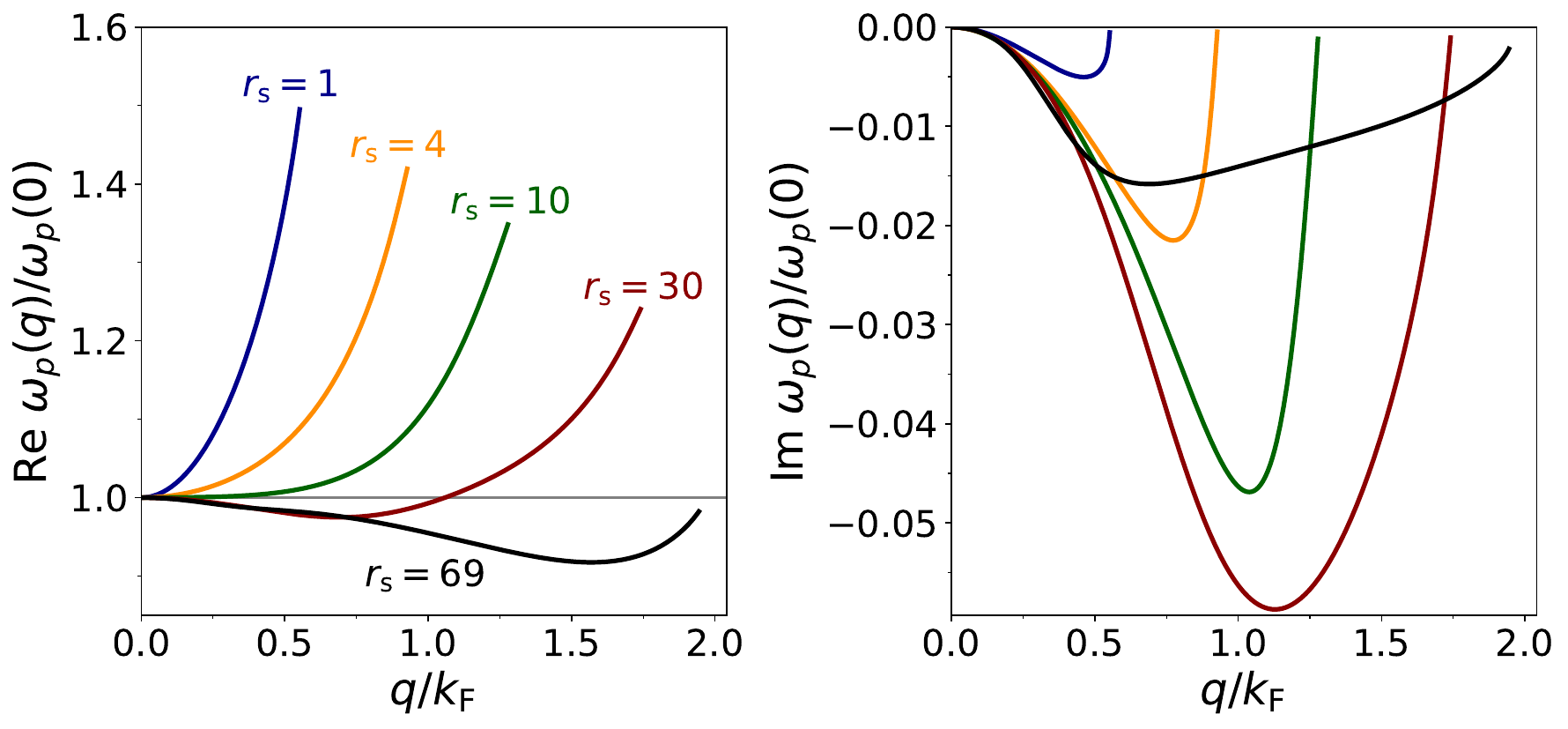}
    \caption{Real (left) and imaginary (right) parts of the rMCP07 plasmon dispersion frequency $\re \omega_p(q)$ such that $|\widetilde{\epsilon}^{\text{rMCP07}}(q,\omega)|<10^{-6}$, with $\widetilde{\epsilon}$ given by Eq. (\ref{eq:df}).}
    \label{fig:TC_plas_disp}
\end{figure*}

The plasmon dispersion curves, plotted in Fig. \ref{fig:TC_plas_disp}, were made by zeroing out the dielectric function at complex frequencies $\omega = u + i v$ (with $u,v$ both real)
\begin{align}
    \widetilde{\epsilon}(q,u + i v) \approx & 1 - \left[\frac{4\pi}{q^2} + \fxc(q,u) - v\frac{\partial \im \fxc(q,u)}{\partial u} \right. \nonumber \\
    & \left. + i v \frac{\partial \re \fxc(q,u)}{\partial u}\right]\chi_0(q,u + i v),
\end{align}
where a low-order Taylor expansion of $\fxc(q,u)$ has been made to analytically continue the kernel to complex frequencies just below the real axis. Without simplification, the Taylor series of $\fxc(q,u)$ would be
\begin{equation}
    \fxc(q,u + i v) \approx \fxc(q,u_0) + (u + iv - u_0)\frac{d \fxc}{d u_0}(q,u_0)
\end{equation}
with $u_0$ a real frequency. In this calculation, we use the Taylor expansion from $u_0=u$ to analytically continue the kernel only to imaginary frequencies. This is more rigorous than the procedure used in Ref. \cite{ruzsinszky2020}, which used a Taylor series about $u_0$, and varied $u$ and $v$. That procedure assumes the low-order Taylor series about $u_0$ also has validity for $u \approx u_0$, which cannot be the case generally.

With that simplification
\begin{align}
    \fxc(q,u + i v) &\approx \fxc(q,u) + i v \frac{\partial \fxc}{\partial u}(q,u) \\
    \fxc(q,u + i v) &\approx \fxc(q,u) + i v \left[\frac{\partial \re \fxc}{\partial u}(q,u) \right. \nonumber \\
    & \left. + i \frac{\partial \im \fxc}{\partial u}(q,u) \right].
\end{align}
As the plasmon frequencies lie just below the real axis, a two-dimensional Newton-Raphson method was used to zero out both components of the dielectric function simultaneously. The Jacobian matrix
\begin{equation}
    \bm{J} = \begin{pmatrix} \frac{\partial \re \widetilde{\epsilon}}{\partial u} & \frac{\partial \re \widetilde{\epsilon}}{\partial v} \\
    \frac{\partial\im \widetilde{\epsilon}}{\partial u} & \frac{\partial \im \widetilde{\epsilon}}{\partial v}
    \end{pmatrix}
\end{equation}
was calculated numerically. Then, given a guess of the plasmon frequency $\omega_{p,j}(q) = u_j + i v_j$, the next guess for the plasmon frequency would be
\begin{equation}
    \begin{pmatrix} u_{j+1} \\ v_{j+1} \end{pmatrix} = (\bm{1} - \bm{J}^{-1})\begin{pmatrix} u_{j} \\ v_{j} \end{pmatrix}.
\end{equation}
The root finding algorithm stopped either when no roots could be found, or when \cite{ruzsinszky2020}
\begin{equation}
    \re \omega_p(q) = \frac{1}{2}q^2 + \kf q,
\end{equation}
indicating that the energies of the plasmon and a particle-hole pair were degenerate. In all cases, we have found that the numerical procedure failed before the particle-hole continuum condition was met.

\begin{figure*}[htbp]
    \centering
    \includegraphics[width=\textwidth]{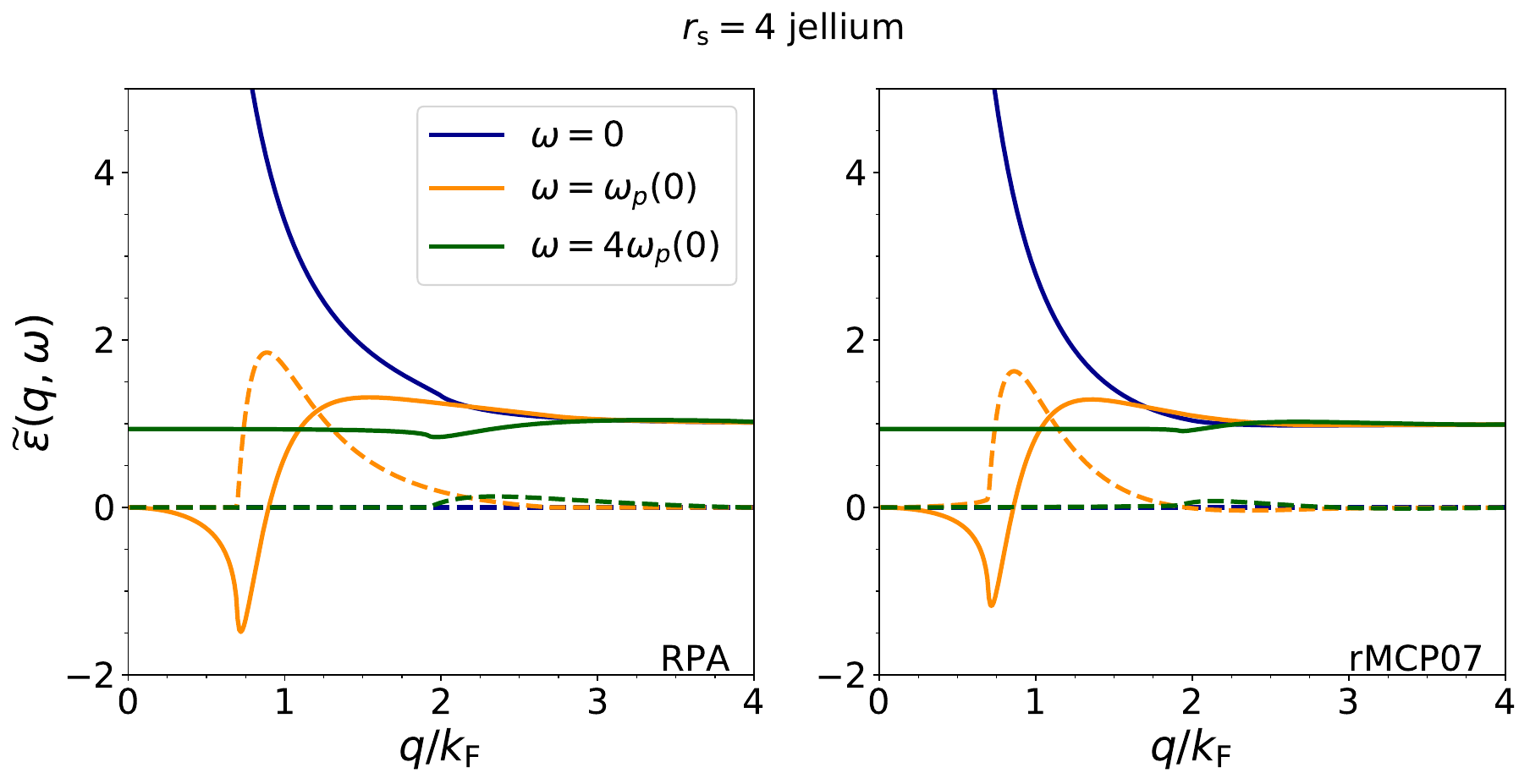}
    \caption{Real (solid) and imaginary (dashed) parts of the RPA (left) and rMCP07 (right) dielectric functions $\widetilde{\epsilon}(q,\omega) = 1 - \left[\frac{4\pi}{q^2} + \fxc(q,\omega)\right]\chi_0(q,\omega)$ for $\omega=0,~\omega_p(0),$ and $4\omega_p(0)$, for $\rs=4$ jellium.}
    \label{fig:TC_epst_rs_4}
\end{figure*}

\begin{figure*}[htbp]
    \centering
    \includegraphics[width=\textwidth]{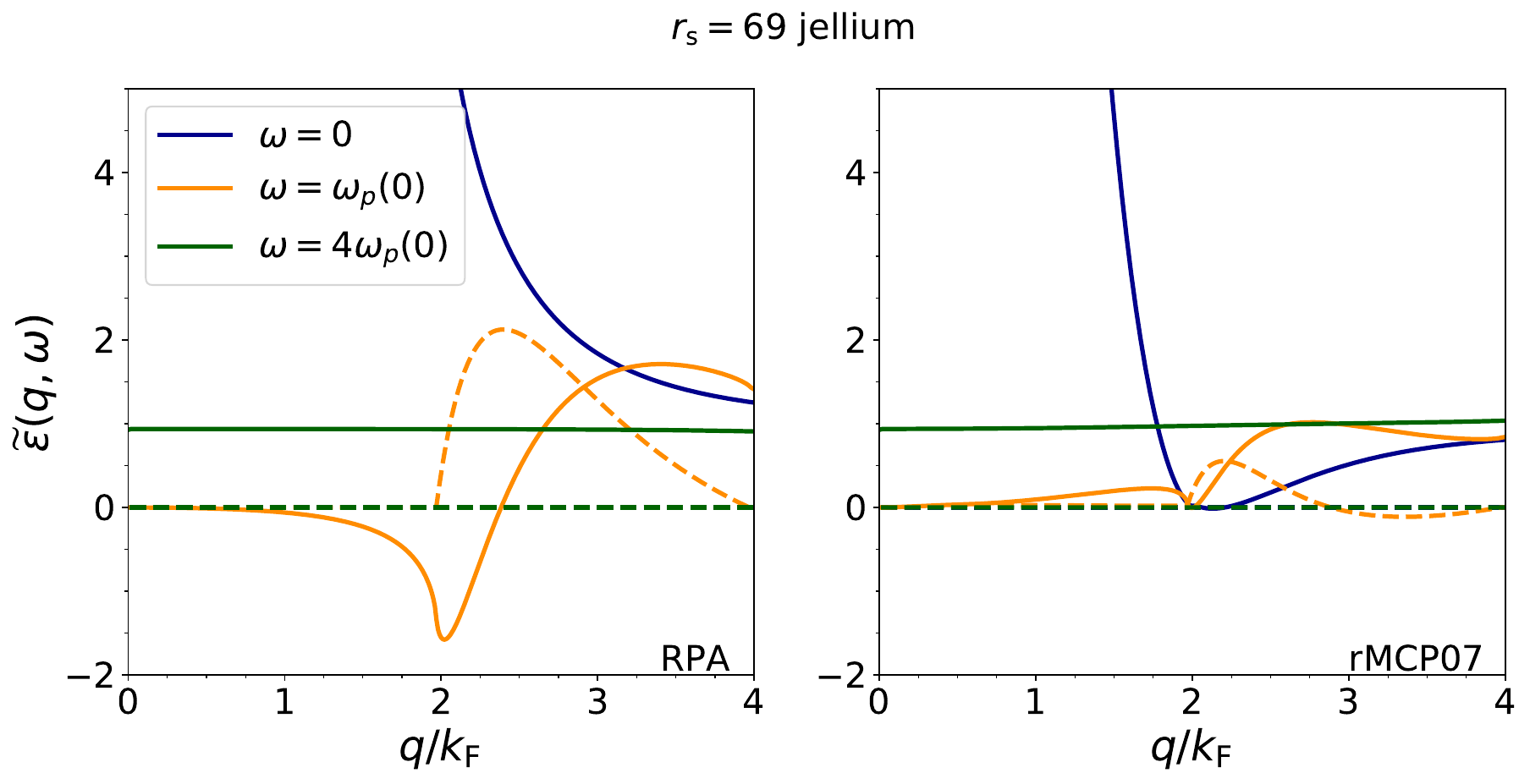}
    \caption{Real (solid) and imaginary (dashed) parts of the RPA (left) and rMCP07 (right) dielectric functions $\widetilde{\epsilon}(q,\omega) = 1 - \left[\frac{4\pi}{q^2} + \fxc(q,\omega)\right]\chi_0(q,\omega)$ for $\omega=0,~\omega_p(0),$ and $4\omega_p(0)$, for $\rs=69$ jellium.}
    \label{fig:TC_epst_rs_69}
\end{figure*}


\section{The jellium structure factor from QMC data}

\begin{figure}[htbp]
    \centering
    \includegraphics[width=\columnwidth]{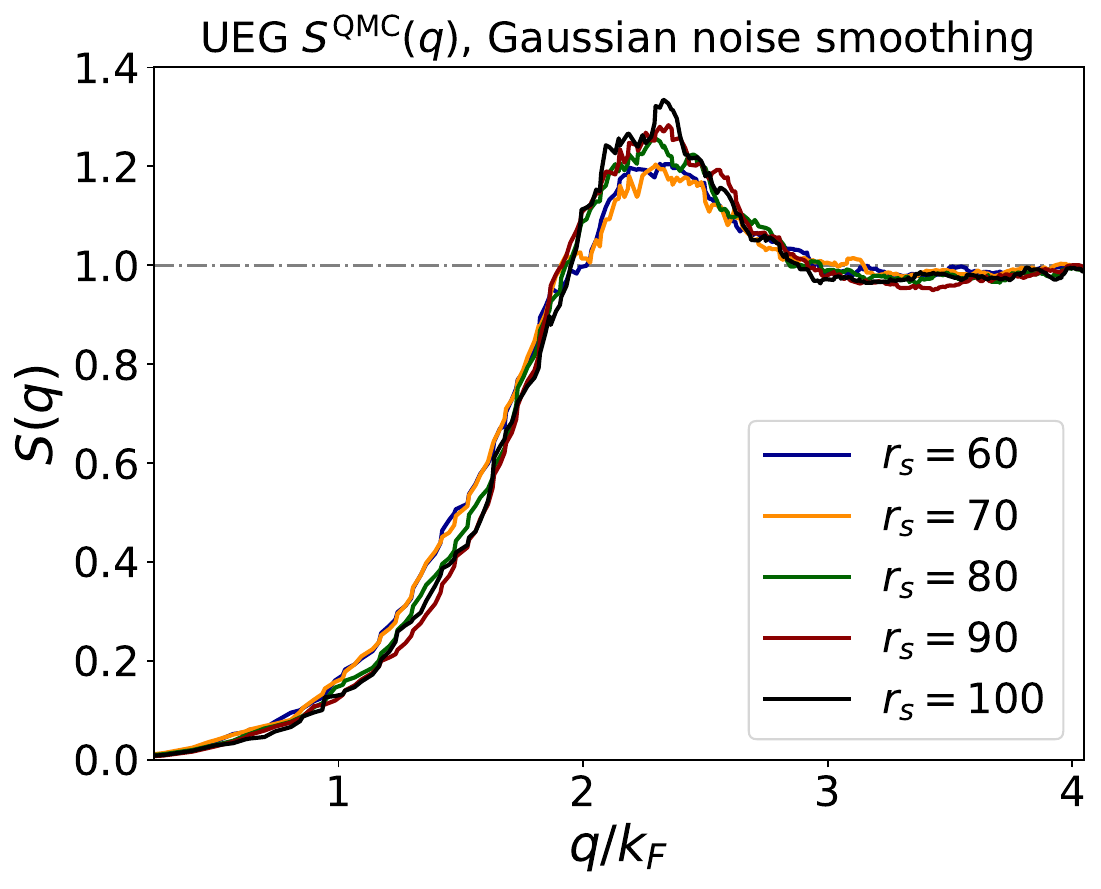}
    \caption{Previously unpublished QMC data of the static structure factor $S(q)$ in jellium \cite{ortiz1999} at lower densities. The data has been smoothed by assuming a Gaussian noise distribution around each point. See the discussion around Eq. (\ref{eq:smoothed_sq}). These results are for the spin-{\it polarized} fluid phase, which was found to be more stable than the spin-{\it unpolarized} fluid phase for $75 \leq \rs \leq 100$ in Ref. \cite{ceperley1980}, and for $60 \leq \rs \leq 100$ in Ref. \cite{ortiz1994}.}
    \label{fig:qmc_dat}
\end{figure}

This section presents previously unpublished QMC data for the static structure factor $S(q)$ of jellium, at lower densities, $\rs\gg 10$.
These results are plotted in Fig. \ref{fig:qmc_dat}, and show that the peak structure in $S(q)$ at intermediate- to low-density jellium is not as pronounced as in MCP07 (Fig. \ref{fig:s_q_comp}).
Details of the QMC computational methods can be found in Refs. \cite{ortiz1994,ortiz1999}.
The structure factors have been computed directly using the Fourier transformed spin-densities $\rho_\sigma(q)$ via $\langle \rho_\sigma(q) \rho_{\sigma'}(-q) \rangle/N$ as described in Refs. \cite{ortiz1994,ortiz1999}.
The calculations used a fixed-node, Jastrow-type trial wavefunction diffusion Monte Carlo (DMC) method, without extrapolation on $S(q)$.
Thus, they are not affected by the limited range of the computed pair distribution function $g(r)$.
More recent improvements in trial wavefunctions would primarily improve the accuracy of extrapolated quantities, but not quantities computed directly [like $S(q)$].
Improvements in techniques, like the backflow method of Ref. \cite{holzmann2020}, would likely not change the location of the wavefunction nodes in a fixed-node DMC calculation.
Further, the results presented here are smoothed (the method is described below).
Therefore, we do not expect the qualitative shapes of the structure factors presented here to change substantially when computed using more recent DMC methods.
An analytic parameterization of the structure factor at high densities $\rs \leq 10$ is given in Ref. \cite{gori2000}.

Note that the data in Fig. \ref{fig:qmc_dat} has been smoothed in the following manner, which we call Guassian noise smoothing. Suppose we sample $S(q)$ at $M$ points $q_0,q_1,...,q_M$, and consider the value of $S(q_i)$ to be correlated to its $2N$-nearest neighbors, at most (by virtue of smoothness). Let $N_L \equiv \max(0,i-N)$ and $N_U\equiv \min(M,i+N)$. Then the smoothed $\widetilde{S}(q_i)$ is given by
\begin{align}
    \widetilde{S}(q_i) &= W^{-1}\sum_{j=N_L}^{N_U}S(q_j)\exp\left\{\frac{[S(q_j)-\mu_i]^2}{\sigma_i}\right\} \label{eq:smoothed_sq}\\
    W &= \sum_{j=N_L}^{N_U}\exp\left\{\frac{[S(q_j)-\mu_i]^2}{\sigma_i}\right\}
\end{align}
for $i=0,1,...,M$, where
\begin{align}
    \mu_i &=\frac{1}{N_U - N_L+1}\sum_{j=N_U}^{N_L} S(q_j) \\
    \sigma_i &=\frac{1}{N_U - N_L+1}\sum_{j=N_U}^{N_L} S(q_j)^2 - \mu^2.
\end{align}
For $q/\kf<1$, $N = 1$, and for $q/\kf \geq 1$, $N=4$. These values were chosen to make a reasonable compromise between data fidelity and readability. The limit $S(q\to 0) \to 0$ is lost when $N$ is increased beyond 1 in this range. Conversely, the raw data (available on the code repository) was too oscillatory near the peak in each curve to be easily interpreted, and thus a larger value of $N$ was needed to smooth the larger, likely unrealistic oscillations. However, increasing $N$ beyond 4 was found to break the limit $S(q\to \infty)\to 1$.

This method of data smoothing is similar to data binning, but with a generalized weight function. Data binning would replace Eq. (\ref{eq:smoothed_sq}) with a simple average,
\begin{equation}
    \widetilde{S}_{\text{bin}}(q_i) = \frac{1}{N_U - N_L}\sum_{j=N_L}^{N_U}S(q_j),
\end{equation}
a method we also tried. However, a simple binning method resulted in lower data fidelity (i.e., too much loss).

\section{The order of limits issue}

The static $\omega\to 0$, long-wavelength $q\to 0$ limit of $\fxc(q,\omega)$ appears to be non-unique. As was derived by Gross and Kohn, \cite{gross1985}
\begin{equation}
    \lim_{q \to 0} \left[\lim_{\omega \to 0} \fxc(q,\omega) \right]= \frac{d^2}{dn^2}[n \varepsilon^{\text{LDA}}\sxc(n)]\equiv \fxc^{\text{ALDA}}(\rs),
\end{equation}
from the compressibility sum rule, where $\varepsilon^{\text{LDA}}\sxc(n)$ is the LDA exchange-correlation energy per electron in jellium. However, as was shown by Conti and Vignale \cite{conti1999}, in the reverse limit
\begin{equation}
    \lim_{\omega \to 0} \left[\lim_{q \to 0} \fxc(q,\omega)\right] = \fxc^{\text{ALDA}}(\rs) + \frac{4}{3}\frac{\mu\sxc(\rs)}{n^2},
\end{equation}
where $\mu\sxc(\rs)$ is the XC shear modulus of bulk jellium. Clearly, both limits agree when $\mu\sxc(\rs) = 0$, however it is unclear what the physical consequences of this assumption would be; the excitation energies of atoms are not described optimally by $\fxc^{\text{ALDA}}$, nor a longitudinal $\fxc(\omega)$ with $\mu\sxc=0$, nor with $|\mu\sxc(\rs)|>0$ \cite{ullrich2004}.

Within time-dependent current-density functional theory \cite{vignale1996}, there exist two kernels in the linear response regime: a longitudinal kernel $\fxc^{\text{L}}$ that is identified with the scalar $\fxc$ of TD-DFT, and a transverse XC kernel $\fxc^{\text{T}}$. In this framework, \cite{conti1999}
\begin{equation}
    \lim_{\omega \to 0} \left[ \lim_{q \to 0} \fxc^{\text{T}}(q,\omega)\right] = \frac{\mu\sxc(\rs)}{n^2}.
\end{equation}
Thus even when $\mu\sxc(\rs)$ is set to zero, an approximation for $\fxc^{\text{T}}(q,\omega)$ can estimate the value of $\mu\sxc(\rs)$. At present, reliable estimates exist only in a limited range of metallic densities \cite{nifosi1998,qian2002}, however $\mu\sxc(\rs)/n^2 \ll |\fxc^{\text{ALDA}}(\rs)|$.

We wish to compare the dynamic GKI kernel with the (longitudinal) dynamic kernel of Qian and Vignale (QV) \cite{qian2002}. The GKI kernel recovers the order of limits $q\to 0$ then $\omega \to 0$, whereas the QV kernel recovers the opposite order of limits. Moreover, the QV kernel promises a more correct treatment of two-plasmon excitations \cite{qian2002} by using a GKI-like frequency interpolation plus a Gaussian correction,
\begin{align}
    \im \fxc(\omega) = -\frac{2\omega_p(0)}{n}&\left\{ \frac{a(\rs)\widetilde{\omega}}{[1 + b(\rs) \widetilde{\omega}^2]^{5/4}} \right. \nonumber \\
    & \left. + \widetilde{\omega}^3\exp\left[ -\frac{(|\widetilde{\omega}|-\Omega(\rs))^2}{\Gamma(\rs)}\right] \right\}, \label{eq:qv_fxc}
\end{align}
where $\widetilde{\omega}=\omega/[2\omega_p(0)]$ and $\omega_p(0)=\sqrt{4\pi n}$ is the semi-classical plasmon frequency. The parameters $a(\rs),~b(\rs),~\Gamma(\rs),$ and $\Omega(\rs)$ are constrained by a set of equations. There are solutions for $a(\rs)$ and $b(\rs)$ for all $\rs$, however there are no solutions for $\Gamma(\rs)$ and $\Omega(\rs) = 1 - 3 \Gamma(\rs)/2$ above a critical $r_{\mathrm{s,c}}$.

Just like the GKI kernel, the QV kernel requires ALDA input; it also requires input for $\mu\sxc(\rs)$ at arbitrary $\rs$. Equation 11 of Ref. \cite{nepal2021} parametrized $\mu\sxc(\rs)$
\begin{equation}
    \frac{\mu\sxc(\rs)}{n} = \frac{a}{\rs} + (b-a)\frac{\rs}{\rs^2 + c}, \label{eq:mu_xc}
\end{equation}
with $a = 0.031152$, $b=0.011985$, and $c=2.267455$; we will use their parametrization here. (Ref. \cite{conti1999} presented a similar fit in Eq. 4.9 of their work, but their parameters appear to be in significant error.) The value of $r_{\mathrm{s,c}}$ above which no solutions exist for $\Gamma(\rs)$ and $\Omega(\rs)$ will depend on the particular $\fxc^{\text{ALDA}}$ and $\mu\sxc(\rs)$ used (PW92 in our case); if $\mu\sxc(\rs)=0$ for all $\rs$, then $r_{\mathrm{s,c}}\approx 45.2$, whereas if Eq. (\ref{eq:mu_xc}) is used, $r_{\mathrm{s,c}}\approx 56.2$.

\begin{figure}
    \centering
    \includegraphics[width=\columnwidth]{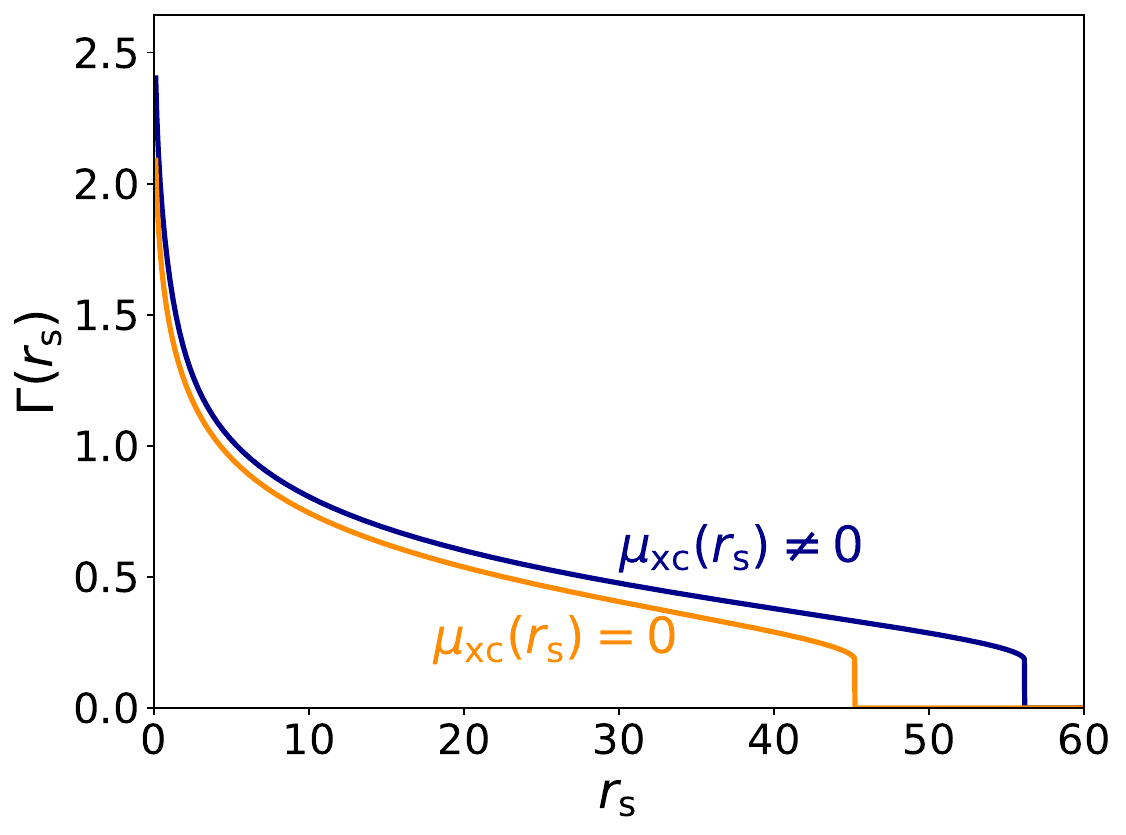}
    \caption{The $\Gamma(\rs)$ parameter in the dynamic, long-wavelength Qian and Vignale \cite{qian2002} kernel. Above a critical $\rs$, no solutions for $\Gamma(\rs)$ can be found consistent with the constraints placed on the kernel. Above this value, we have set $\Gamma=0$; the transition is abrupt, and dependent upon the ALDA used, as well as the XC shear modulus.}
    \label{fig:qv_rsc}
\end{figure}

For all $\rs > r_{\mathrm{s,c}} $, we are forced to set $\Gamma = \epsilon$, where ideally $\epsilon=0$, but in practice $\epsilon = 10^{-14}$. This yields essentially a double-delta function resonance at $\omega = \pm 2\omega_p(0)$, signaling onset of a two-plasmon excitation. As seen in Fig. \ref{fig:qv_rsc}, the value of $\Gamma(\rs)$ abruptly falls to zero for $\rs > r_{\mathrm{s,c}}$.

\begin{figure}
    \centering
    \includegraphics[width=\columnwidth]{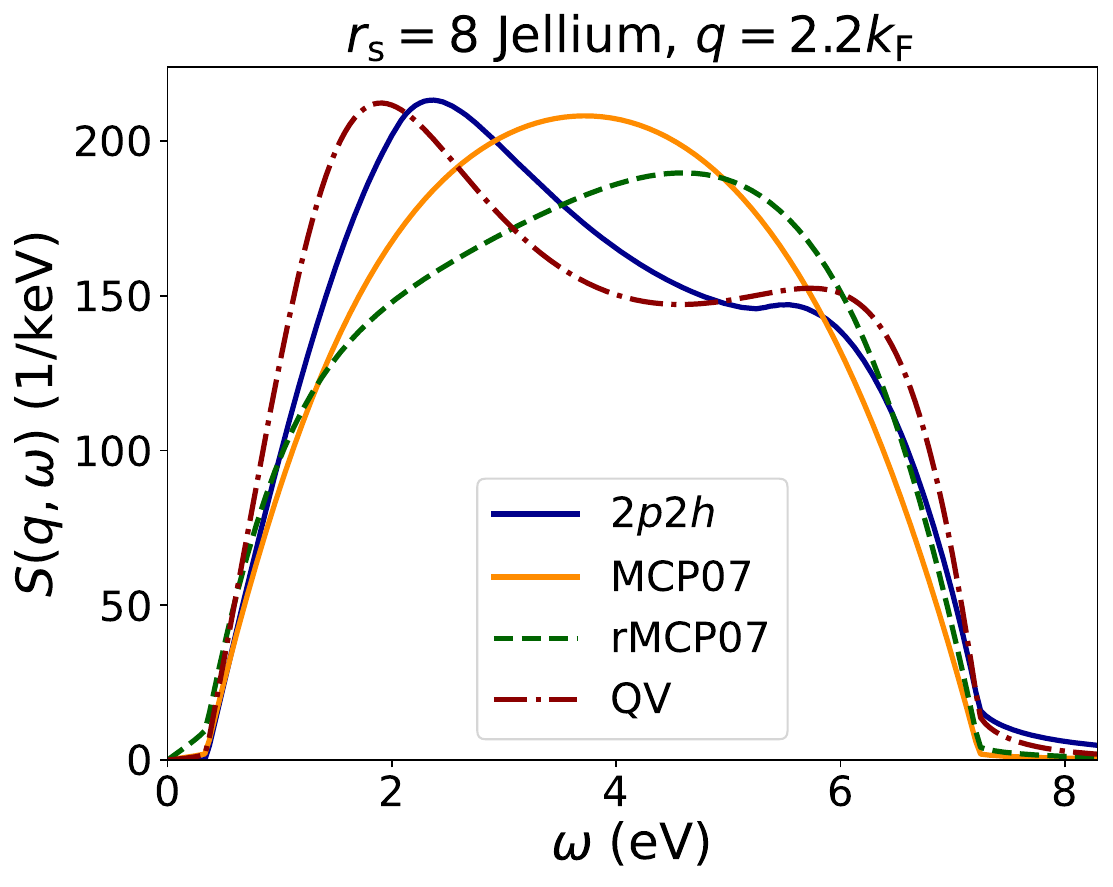}
    \caption{Comparison of the dynamic structure factor $S(q=2.2\kf,\omega)$ for various model kernels for $\rs=8$ jellium, analogous to Fig. 2 of Ref. \cite{panholzer2018}. The ghost exciton can be observed as a double-peak structure in the $2p2h$ data and QV kernel only. }
    \label{fig:ghost_exciton}
\end{figure}

The QV kernel is able to capture excitonic excitations, due to the Gaussian term in Eq. (\ref{eq:qv_fxc}), which reduces to a delta-function resonance at low densities. Figure \ref{fig:ghost_exciton} shows that the QV kernel predicts the emergence of a ``ghost exciton'' in intermediate density jellium.

For reasons that have been described in the Introduction, we have not fitted a QV-MCP07 kernel, where the frequency-dependence of the GKI kernel is replaced by that of the QV kernel. Whereas we can easily deduce a parameterization of the real part of the GKI kernel that is independent of $\rs$, and thus also a reasonable parameterization of its continuation to imaginary frequencies, a similar procedure cannot be done for the QV kernel. The GKI-like part of the QV kernel can be expressed using Eq. (\ref{eq:new_hx}), however the real part of the Gaussian term cannot be expressed in an $\rs$-independent form, nor can the real part be computed analytically. We found that a low-order Taylor expansion of the real part of the kernel rapidly breaks down for $\omega/\omega_p(0) \ll 1$, and is thus not useful in a Pad\'e-like approximant.

The rMCP07 fitting involves only a three-dimensional integration that can be rapidly expedited using parallel computation. The QV-MCP07 fitting would involve a five-dimensional numeric integration at each value of the interaction-strength--scaled frequency, which cannot be as easily parallelized.

\section{Ultranonlocality coefficient \label{sec:app_unonloc}}

\begin{figure}[htbp]
    \centering
    \includegraphics[width=\columnwidth]{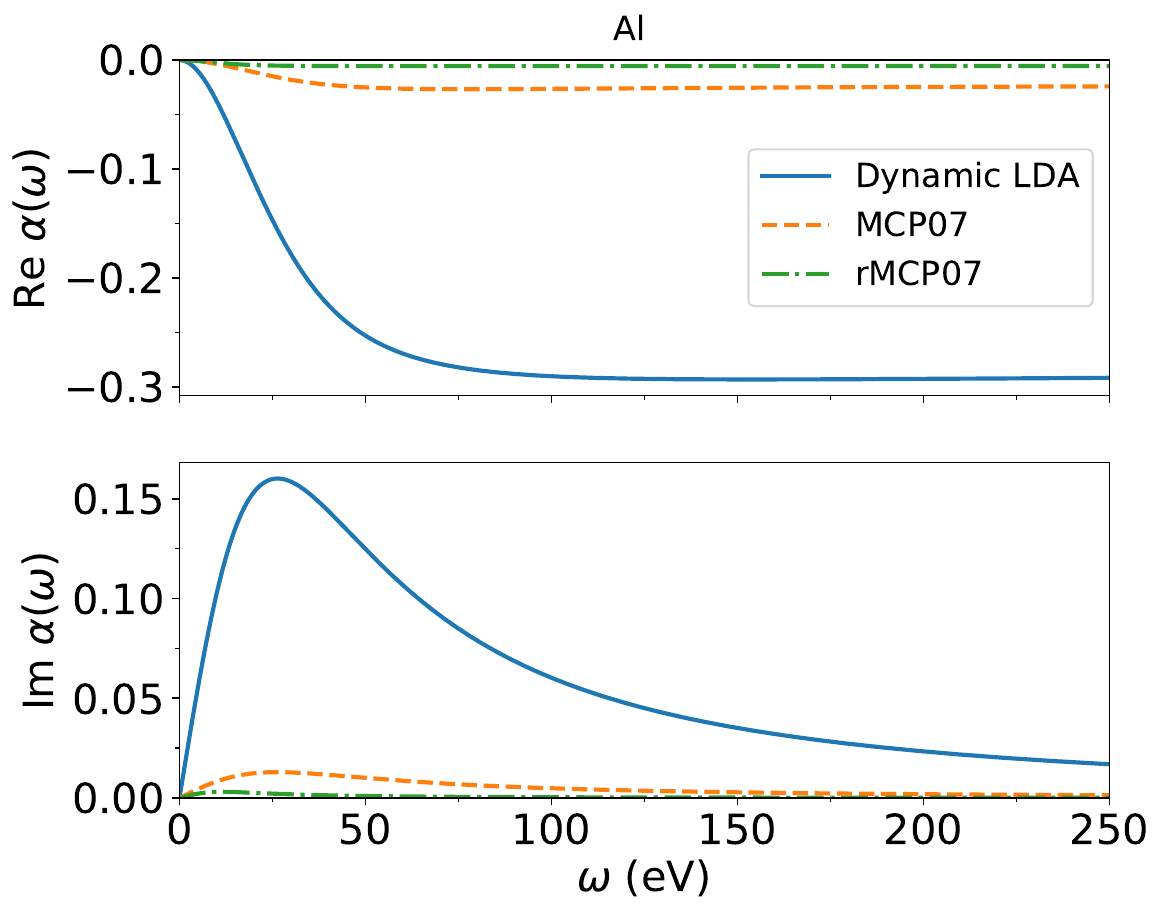}
    \caption{The ultranonlocality coefficient $\alpha(\omega)$ in face-centered cubic Al, using the same pseudopotential density as was used in Ref. \cite{nepal2021}. The dynamic LDA refers to the GKI frequency-dependent kernel, but using Eq. (\ref{eq:new_hx}) to model the real part of $\fxc(q=0,\omega)$, and with PW92 for the ALDA.}
    \label{fig:unonloc_al}
\end{figure}

\begin{figure}[htbp]
    \centering
    \includegraphics[width=\columnwidth]{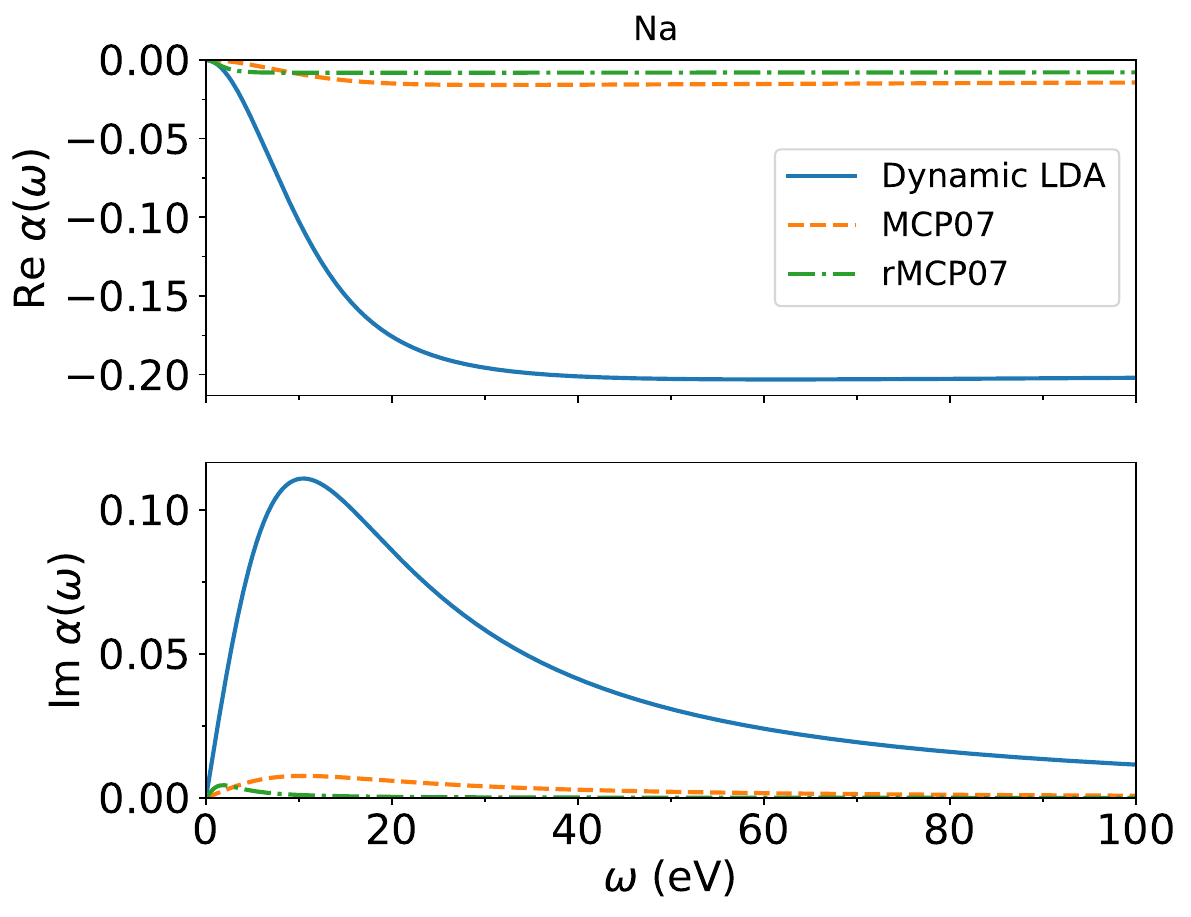}
    \caption{The ultranonlocality coefficient $\alpha(\omega)$ in body-centered cubic Na, using the same pseudopotential density as was used in Ref. \cite{nepal2021}.}
    \label{fig:unonloc_na}
\end{figure}

As in Ref. \cite{nepal2021}, this section computes the ultranonlocality coefficient $\alpha(\omega)$ \cite{nazarov2009}
\begin{equation}
    \lim_{|\bm{q}|\to 0} \fxc(\bm{q},\bm{q},\omega) = -\frac{4\pi\alpha(\omega)}{q^2}.
\end{equation}
$\alpha(\omega)$ is the frequency-dependent strength of the long-range part of $\fxc$. $\alpha(\omega)$ vanishes for a uniform density.
For a weakly-inhomogeneous density, such as that of a real simple metal, we have computed
$\alpha(\omega)$ by the formula of Ref. \cite{nazarov2009}.
This $\alpha(\omega)$ is plotted in Figs. \ref{fig:unonloc_al} and \ref{fig:unonloc_na}.
For an insulator, $\alpha(\omega)$ has significant effects on optical absorption.

\section{Density fluctuations \label{sec:dens_fluc}}

This section deals with frequency moments of the dynamic structure factor
\begin{equation}
    M_\omega^k(q) = \int_0^\infty S(q,\omega)\omega^k d\omega,
    \quad k=0,1,2,...
\end{equation}
Reference \cite{perdew2021} suggested that the following frequency moments, weighted by the static structure factor $M_\omega^0(q)=S(q)$,
\begin{align}
    \langle \omega_p(q) \rangle &= \frac{M_\omega^1(q)}{M_\omega^0(q)} \\
    \langle \Delta \omega_p(q) \rangle &= \left[ \frac{M_\omega^2(q)}{M_\omega^0(q)} -  \langle \omega_p(q) \rangle^2\right]^{1/2}
\end{align}
could describe the average and standard deviation in the frequency of a density fluctuation, respectively. Their analysis demonstrated that, in low density jellium, the average frequency of a density fluctuation abruptly drops towards zero for $q \approx 2 \kf$. This would suggest the emergence of a charge-density wave at low density within Anderson's \cite{anderson1972} interpretation of symmetry breaking: Fluctuations in the density of a large number of electrons can abruptly freeze, signaling the onset of an observable symmetry broken phase that would not be observable in a system of few electrons.

\begin{figure}
    \centering
    \includegraphics[width=\columnwidth]{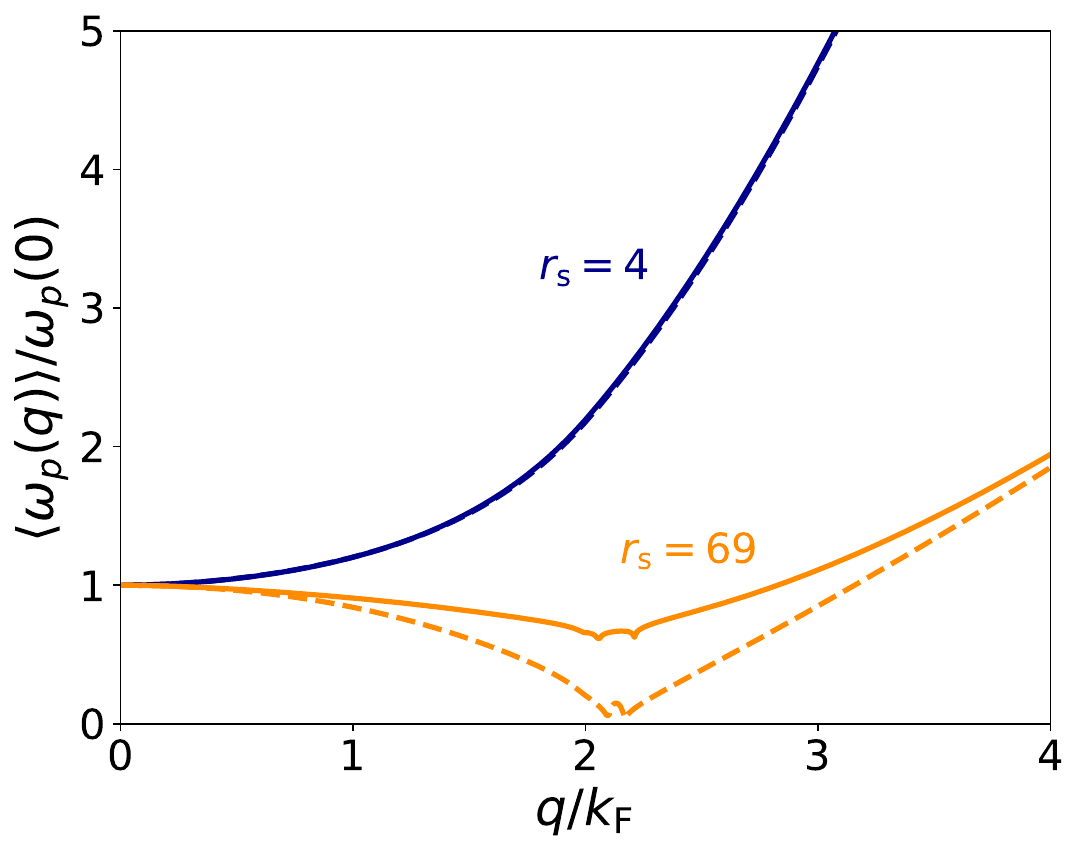}
    \caption{Average density fluctuation $\langle \omega_p(q) \rangle$ in bulk jellium for the MCP07 (dashed) and rMCP07 (solid) kernels. The curves essentially coincide at $\rs=4$, but differ sharply at $\rs=69$.}
    \label{fig:avg_dens_fluc}
\end{figure}

\begin{figure}
    \centering
    \includegraphics[width=\columnwidth]{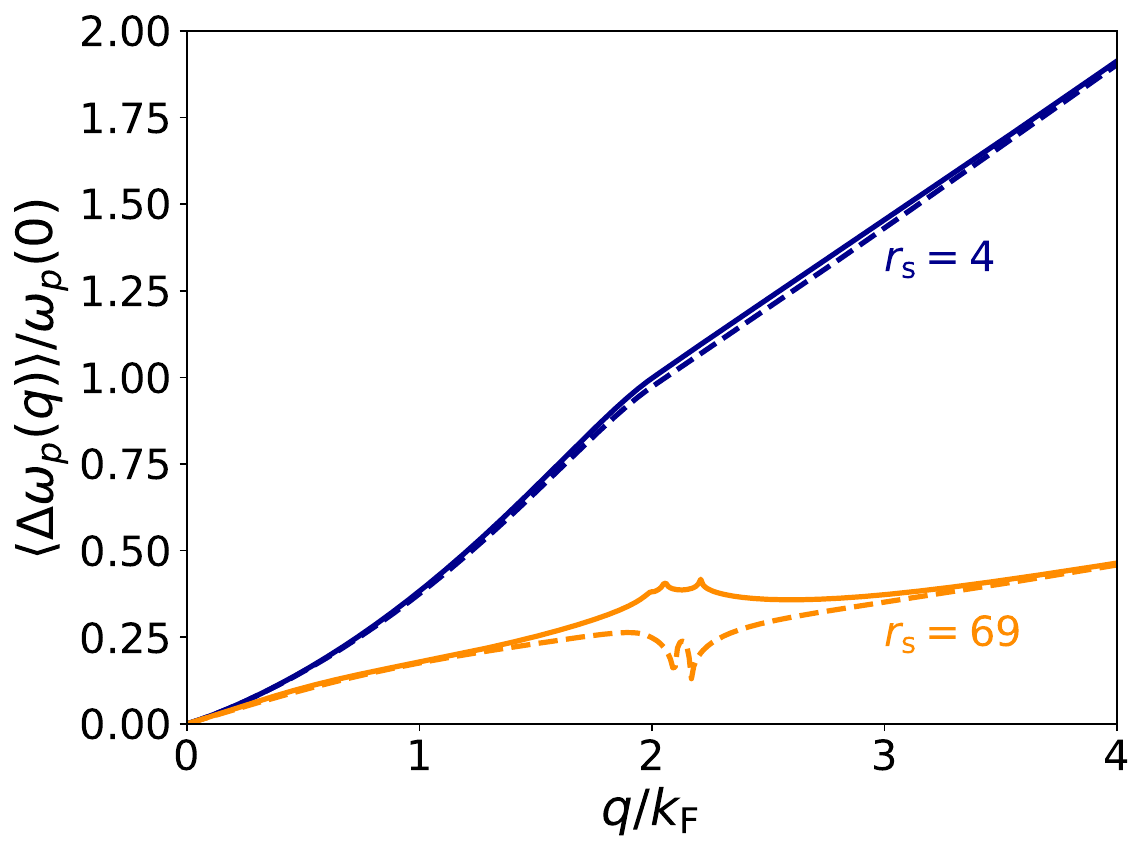}
    \caption{Standard deviation in the density fluctuation $\langle \Delta \omega_p(q) \rangle$ in bulk jellium for the MCP07 (dashed) and rMCP07 (solid) kernels. The curves mostly coincide at $\rs=4$ and differ in slope and concavity at $\rs=69$.}
    \label{fig:stddev_dens_fluc}
\end{figure}

This behavior can be observed in Fig. \ref{fig:avg_dens_fluc} for the MCP07 kernel. Interestingly, the rMCP07 value of $\langle \omega_p(q) \rangle$ does not drop to zero at $\rs = 69$. Figure \ref{fig:stddev_dens_fluc} displays $\langle \Delta \omega_p(q) \rangle$.

Therefore, the rMCP07 kernel does not describe the low-density fluctuations of jellium well, at least within our first interpretation \cite{perdew2021} of Anderson's theory of symmetry breaking. It seems likely to us that the  spectral weight at or near the critical density and wavevector should drop to a small frequency, but not to zero frequency.

This behavior of rMCP07 is due to the scaling function $p(q,\rs)$ of Eq. (\ref{eq:pscl}). $p(q,\rs)$ decreases the rate at which $\fxc(0,\Omega)$ approaches its infinite frequency limit for $\rs < C \approx 4.35$ bohr. Conversely, for $\rs > C$, $\fxc(0,\Omega)$ more rapidly approaches its infinite frequency limit. This behavior, while seemingly necessary for the recovery of accurate correlation energies, introduces a questionable zero to the real part of the effective dielectric function $\widetilde{\varepsilon}$ at nonzero frequency, as seen in Fig. \ref{fig:TC_epst_rs_69}, and thus a questionable pole into $S(q,\omega)$ at the same nonzero frequency.

This behavior can also be tied to the spectral function $S(q)$ at lower densities. Consider Fig. \ref{fig:qmc_dat}, which plots $S^{\text{QMC}}(q)$ for the spin-polarized fluid phase. Although $S^{\text{rMCP07}}(q)$, plotted in Fig. \ref{fig:s_q_new}, and $S^{\text{MCP07}}(q)$, plotted in Fig. \ref{fig:s_q_comp}, are for the spin-unpolarized fluid phase, it is clear that rMCP07 gives a more realistic description of the ground state $S(q)$ than does MCP07. This is because the peak structure in $S^{\text{MCP07}}(q)$ is softened dramatically in $S^{\text{rMCP07}}(q)$. This softening is also observed in Fig. \ref{fig:avg_dens_fluc}, where the average frequency of a plasmon is much smoother in rMCP07, never dropping to zero frequency.

\section{Note on methods employed here}

All calculations were performed using libraries written by the authors in Python 3 and Fortran 90 \cite{code_repo}. The numeric methods employed are varied, so we mention only a few specific ones here. Kramers-Kronig and Cauchy principal value integrals were evaluated using adaptive Gauss-Kronrod quadrature. Multi-dimensional integrations, and frequency moment integrations, were performed with Gauss-Legendre quadrature grids along each axis. For details of the frequency moment calculation and the Gauss-Kronrod integrator, we refer the reader to the Supporting Information of Ref. \cite{perdew2021}. For calculation of the right-hand side of Eq. (\ref{eq:m3_sr}) (third moment sum rule), the static structure factor was tabulated at each value of $\rs$ and interpolated using cubic splines.

The GKI kernel parameters ($c_i$ and $k_i$) were fitted in two steps: initial parameters were determined by a least squares search, and these were further refined by a grid search. The rMCP07 parameters ($A$, $B$, $C$, and $D$) were determined in a similar fashion, however the initial fit was determined by a Nelder-Mead simplex algorithm.

Calculation of the critical wavevector for onset of a static charge density wave was performed using a bisection root finding algorithm. The plasmon dispersion curves were generated using a Newton-Raphson root finding method; a full discussion is given in Appendix A. For a discussion of the ultranonlocality coefficient calculation, we refer the reader to Ref. \cite{nepal2021}.

\end{document}